\documentclass[mathleft
]{an}
\usepackage{graphicx}

\usepackage{times}
\begin{document}

\Pagespan{1}{}
\Yearpublication{2010}%
\Yearsubmission{2010}%
\Month{07}%
\Volume{999}%
\Issue{88}%

\title{Measuring the mass accretion rates of Herbig Ae/Be stars\\with X-shooter
\thanks{Based on observations obtained at the European Southern Observatory, Paranal, Chile
(ESO programme 385.C-0131(A)).}}

\date{Received date / Accepted date}

\author{{M.~A.~Pogodin\inst{1,2}\fnmsep\thanks{Corresponding author: \email{pogodin@gao.spb.ru}}}
\and
S.~Hubrig\inst{3}
\and
R.~V.~Yudin\inst{1,2}
\and
M.~Sch\"oller\inst{4}
\and
J.~F.~Gonz\'alez\inst{5}
\and
B.~Stelzer\inst{6}
}

\titlerunning{Measuring the mass accretion rates of Herbig Ae/Be stars with X-shooter}
\authorrunning{M.~A.~Pogodin et al.}

\institute{
Central Astronomical Observatory at Pulkovo, St.~Petersburg 196140, Russia
\and
Isaac Newton Institute of Chile, Saint-Petersburg Branch, Russia
\and
Leibniz-Institut f\"ur Astrophysik Potsdam, An der Sternwarte 16, 14482 Potsdam, Germany
\and
European Southern Observatory, Karl-Schwarzschild-Str.~2, 85748 Garching, Germany
\and
Instituto de Ciencias Astronomicas, de la Tierra, y del Espacio (ICATE), 5400, San Juan, Argentina
\and
INAF-Osservatorio Astronomico di Palermo, Piazza del Parlamento 1, 90134 Palermo, Italy
}



\abstract{
We present the results of our observations of eight magnetic Herbig Ae/Be stars
obtained with the X-shooter spectrograph mounted on UT2 at the VLT. X-shooter
provides a simultaneous, medium-resolution and high-sensitivity spectrum over the
entire wavelength range from 300 to 2500\,nm. We estimate the mass accretion rates
(${\dot{M}}_{\rm acc}$) of the targets from 13 different spectral diagnostics using
empiric calibrations derived previously for T\,Tauri-type stars and brown dwarfs.
We have estimated the mass
accretion rates of our targets, which range from $2\times10^{-9}$ to
$2\times10^{-7}$\,$M_{\sun}$/yr. Furthermore, we have found accretion
rate variability with amplitudes of
0.10--0.40\,dex taking place on time scales from one day to tens of
days. Additional future night-to-night observations need to be
carried out to investigate the character of ${\dot{M}}_{\rm acc}$
variability in details. Our study shows
that the majority of the calibrational relations can be applied to Herbig
Ae/Be stars, but several of them need to be re-calibrated on the basis of new
spectral data for a larger number of Herbig Ae/Be stars.
}

\maketitle

\section{Introduction}
\label{sect:intro}

Herbig Ae/Be stars (HAeBes) are pre-main sequence (PMS) objects of
intermediate mass approximately from 2 to 10\,$M_{\sun}$
(Herbig \cite{Herbig.1960}; Finkenzeller \& Mundt \cite{FinkenzellerMundt1984}).
They range from spectral class F2--O9
and are surrounded by relict dust/gas accretion disks, reflecting
themselves in the form of an IR excess as a result of thermal
emission of circumstellar (CS) dust. For a number of stars, the disks
can also be directly detected on images
(e.g.\ Grady et al.\ \cite{Grady2005}). The CS environment of HAeBes possesses a
complex structure. Gaseous streams are accreted from the equatorial disk onto the
star and a  stellar wind zone exists at higher latitudes.  On the other hand, the
disk wind, the so-called magnetocentrifugal wind, carries the excess of angular
momentum away
(e.g.\ Garcia et al.\ \cite{Garcia2001}). CS contribution to the stellar atmospheric spectrum
provides information on physical processes in the envelope and on its interaction
with the star. The character of this interaction is reasonably well studied for PMS
objects with masses of about one solar mass and less (T\,Tauri-type stars or TTS).
These objects have significant (several kG) dipole-type magnetic fields. The star/CS
interaction in TTS is described in detail by the  model of magnetospheric accretion
(MA; e.g.\ Bouvier et al.\ \cite{Bouvier2007}, and references therein). According to the MA-model,
the field truncates the disk where the ram pressure of the accreted matter is
balanced by pressure from the stellar magnetic field. At the truncation radius the
accreted material is channeled into funnel flows and ballistically falls onto the
stellar surface. A situation that is less clear for HAeBes. Their interiors are
mostly radiative (see e.g.\ Fig.~8 in Hubrig et al.\ \cite{Hubrig2009}). However, convection is
needed to maintain a stellar dynamo that generates  a strong magnetic field.
Nevertheless, most late-type HAeBes with spectral classes later than B9 (HAes), have
possibly evolved from the so-called intermediate mass T\,Tauri stars (IMTTS) of K--G spectral
types (see e.g. Palla \& Stahler \cite{PallaStahler1993}) possessing sub-atmospheric convective zones and, as a
result, strong magnetic fields. It is likely that these fields do not dissipate
completely during the PMS evolutionary stage. Additionally, it is possible that
magnetic fields of HAeBes are generated by differential rotation of the star or
originate in dynamos within the CS disk (Tout \& Pringle \cite{ToutPringle1995}; Tout \& Pringle \cite{ToutPringle1996}).
Recent direct measurements of HAeBe magnetic fields using spectropolarimetry
(Wade et al.\ \cite{Wade2007}; Hubrig et al.\ \cite{Hubrig2007}, \cite{Hubrig2009}, \cite{Hubrig2011a}, and references therein) have allowed to
detect longitudinal magnetic fields of hundreds of Gauss in a number of HAes.
Stars for which so far no magnetic field could be detected may still have a
field but below the current sensitivity limits.  The inner boundary of their
accretion disks are likely to be located close to the stellar surface.

The existence of magnetically controlled disk accretion in HAes going through polar
funnels was confirmed by Muzerolle et al.\ (\cite{Muzerolle2004}), Mottram et al.\
(\cite{Mottram2007}), and Grady et al.\ (\cite{Grady2010}). In any case, the
magnetospheric radii of HAes with B$\approx$100\,G must be much smaller in
comparison with those of TTS, where magnetic fields are of the order of several kG.
As a result, we can expect the character of the star/CS interaction in HAes to be
different from that of TTS. In our study aim at investigating the general picture of
the star/CS interaction in HAes and at comparing it with that in TTS. One of the
initial goals was also to study the variations of the accretion mass rates over the
magnetic period. However, due to weather or technical constraints the observations
were carried out mostly for targets with single magnetic field measurements,  i.e.\
for targets with unknown magnetic phase curves. Clearly, an analysis of the temporal
behaviour of different mass accretion rate indicators over the magnetic/rotation
periods would provide fundamental clues on how magnetospheric accretion works. The
targets for which magnetic/rotation periods have been determined in our previous
studies include HD\,97048 and HD\,101412 with only one X-shooter observations,
respectively, HD\,150193 with two observations, and HD\,176386 with four
observations (Hubrig et al.\ \cite{Hubrig2011b}, \cite{Hubrig2011a}).

One of the most important parameters characterizing the properties of the star/CS
interaction is the mass accretion rate ${\dot{M}}_{\rm acc}$. It determines strongly
the disk structure and dynamics (e.g., D'Alessio et al.\ \cite{DAlessio1999}, \cite{DAlessio2001}).
Measurements of ${\dot{M}}_{\rm acc}$ are well-developed for TTS, based on a number
of empirical spectral diagnostics (e.g., Oudmaijer et al.\ \cite{Oudmaijer2011}, Rigliaco et al.\ \cite{Rigliaco2011}, and
references therein).

Three main purposes of our work presented here are:
\begin{enumerate}
\item To determine ${\dot{M}}_{\rm acc}$ in a sample of HAes whose
magnetic fields have been detected (Hubrig et al.\ \cite{Hubrig2009},
\cite{Hubrig2011b}) using calibrations previously obtained for TTS and brown dwarfs
(BDs). Magnetic HAes appear to be most suitable targets for such a study because
their disk accretion process is expected to better fit the MA scenario in comparison
with  non-magnetic stars.
\item To test the applicability of different calibrations for
measuring ${\dot{M}}_{\rm acc}$ of HAes  and to select the best spectroscopic
tracers of accretion necessary to provide accurate accretion rates for systematic
studies of large samples of HAes.
\item To search for variability of ${\dot{M}}_{\rm acc}$ in HAes with
more than one available spectrum. This variability can be twofold: a) a real change
in the accretion process, and b) a modulation of ${\dot{M}}_{\rm acc}$ by the
rotating magnetosphere and a cyclic screening of the stellar limb by local funnel
streams.
\end{enumerate}

\section{Mass accretion rate calibrations}
\label{sect:macc_calibs}

\subsection{Calibrations for TTS and BDs}
\label{sect:calib_ttsbds}

At present, a number of empiric spectral indicators are available for the
determination of ${\dot{M}}_{\rm acc}$ for TTS and BDs.

\subsubsection{Model method} \label{sect:calib_model}

The basic accretion indicator is the additional emission in the continuum
superimposed onto the stellar spectrum and calculated on the basis of MA models for
these objects (e.g., Calvet \& Gullbring \cite{CalvetGullbring1998}, Gullbring et al.\ \cite{Gullbring2000}, Ardila et al.\ \cite{Ardila2002}).
It is assumed that the
overwhelming bulk of this emission originates in funnel flows inside the
magnetosphere and in the shock zone on the stellar surface in a compact region. This
assumption is rather plausible since the emissivity in the continuum is proportional
to $N_{e}^{2}$ (where $N_{e}$ is the electron density), and the magnetospheric
streams consist of the densest CS matter. The gas density in other parts of the
envelope (disk and wind) drops very quickly with distance from the star, and its
contribution to the additional continuum emission is  negligible. In the observed
spectra, this emission is seen as veiling of the atmospheric lines and as an excess
radiation at wavelengths below the Balmer jump. The theoretical models allow to
construct a quantitative relation between ${\dot{M}}_{\rm acc}$ and the veiling plus
UV excess. Yet, veiling and UV excess frequently prove not to be convenient for
these measurements, due to the large extinction in the regions where the targets are
located.

\subsubsection{L - type calibrations} \label{sect:calib_l}

Alternatively, empirical calibrations have been determined between emission lines
and mass accretion.

All emission lines in the spectrum of a PMS object are of multi-component origin.
They are formed not only in the magnetospheric streams, but also in the disk and in
the wind. The source function determining the emissivity in a line drops with the
distance from the star much slower than the emissivity in the continuum (for
example, see Pogodin \cite{Pogodin.1986,Pogodin.1989} for  Balmer  emission lines).
However, according to the MA models, such parameters as the mass of the star, mass
of the disk, ${\dot{M}}_{\rm acc}$ and mass loss rate ${\dot{M}}_{\rm wind}$ are
likely to be interdependent. It has been established that the accretion luminosity
($L_{\rm acc}$) is clearly correlated with luminosity in some emission lines
($L_{\rm line}$). A number of empirical calibrations have been published so
far . The dependencies are approximated by expressions of the type:

\begin{equation}
\log(L_{\rm acc}) = a \log (L_{\rm line}) + b,
\end{equation}

\begin{equation}
{\dot{M}}_{\rm acc} = \frac{L_{\rm acc} R_{*}}{G M_{*}},
\end{equation}

\noindent

where $M_{*}$ and $R_{*}$ are the stellar mass and radius, respectively, and $a$ and
$b$ are coefficients determined separately for each spectral line (e.g.,
Fang et al.\ \cite{Fang2009}, Dahm \cite{Dahm.2008}, Herczeg \& Hillenbrand \cite{HerczegHillenbrand2008},
Gatti et al.\ \cite{Gatti2008}, Natta et al.\ \cite{Natta2004}, Mohanty et al.\ \cite{Mohanty2005}). The calibrations are constructed
on the basis of spectroscopic observations of the chosen line in a representative sample of targets,
whose ${\dot{M}}_{\rm acc}$ values were determined beforehand using the
model method.

\subsubsection{F - type calibrations} \label{sect:calib_f}

Besides of that, another type of empirical calibrations exists that is based on a
direct correlation  between ${\dot{M}}_{\rm acc}$ and the flux in the emission line
($F_{\rm line}$). The dependencies are expressed as:

\begin{equation}
\log({\dot{M}}_{\rm acc}) = c \log (F_{\rm line}) + d,
\end{equation}

\noindent
where $c$ and $d$ are coefficients (e.g.\ Herczeg \& Hillenbrand \cite{HerczegHillenbrand2008}). The F-type
calibrations are constructed analogous to the case of L-type calibrations
where the values of ${\dot{M}}_{\rm acc}$ are derived beforehand using the model
method or from other types of calibration.

\subsubsection{{H$\alpha 10\%$} - indicator}
\label{sect:calib_h}

There is also a separate accretion indicator, the H$\alpha 10\%$. It is defined as
the full width W  of the emission H$\alpha$ profile at the level of $10\%$ of the
maximum intensity  (White \& Basri \cite{WhiteBasri2003}). Natta et al.\ (\cite{Natta2004}) extended this indicator to very
low-mass objects, where ${\dot{M}}_{\rm acc}$ was determined by fitting the observed
H$\alpha$ profiles with predictions of MA models (Muzerolle et al.\ \cite{Muzerolle2001}). The
calibration is expressed in the form:

\begin{equation}
\log({\dot{M}}_{\rm acc}) = c_1 \log (W(H\alpha 10\%)) + d_1,
\end{equation}

\noindent
where $c_{1}$ and $d_{1}$ are coefficients and W(H$\alpha 10\%$) is given in (km/s)
(Natta et al.\ \cite{Natta2004})

\subsection{Calibrations for HAeBes}
\label{sect:calib_haebes}

\begin{figure}
\centering
\includegraphics[width=0.45\textwidth]{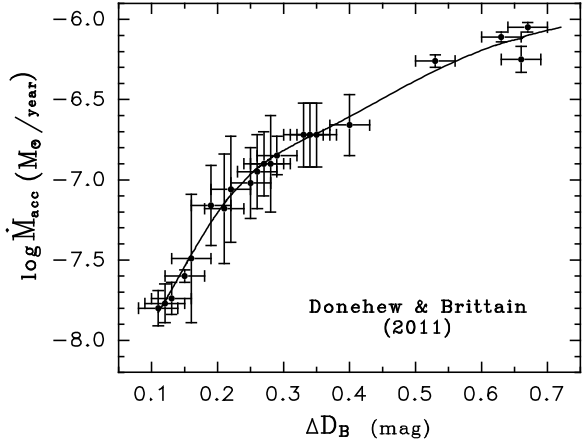}
\caption{
Relationship between $\Delta D_{B}$ and ${\dot{M}}_{\rm acc}$ constructed
on the base of data extracted from Table~3 in Donehew \& Brittain (\cite{DonehewBrittain2011}) and
fitted by a polynomial.
}
\label{f1}
\end{figure}

At present, calibrations constructed specifically for HAeBes are small in number.

Donehew \& Brittain (\cite{DonehewBrittain2011}) derived ${\dot{M}}_{\rm acc}$ for a considerable sample of HAeBes
using the model method,  where the modeling was applied to the emission Balmer jump
$\Delta D_{B}$. They calculated ${\dot{M}}_{\rm acc}$ in dependence of the emission
Balmer jump for more than 30 HAeBes using the models and methods developed by
Calvet \& Gullbring (\cite{CalvetGullbring1998}), Gullbring et al.\ (\cite{Gullbring2000}), and Muzerolle et al.\ (\cite{Muzerolle2004}).
The model of Muzerolle et al.\ (\cite{Muzerolle2004}) takes into account the main physical processes predicted by the MA
model such as: a ballistical infall of material from the disk onto the stellar
surface along accretion columns, heating of the photosphere due to its shock
interaction with the accretion streams and releasing soft X-rays that are absorbed
by the surrounding material. This material then emits optical and UV radiation as it
thermalizes. In the model, the flux from accretion is calculated for three different
regions: the shock region, the heated photosphere, and the accretion columns, to get
the overall flux from accretion, which is a function of $M_{*}$, $R_{*}$, and
${\dot{M}}_{\rm acc}$. The resulting spectral energy distribution (SED) including
both the accretion and the stellar flux is used to calculate the expected emission
Balmer jump $\Delta D_{B}$ for different ${\dot{M}}_{\rm acc}$ and stars of
different spectral types. The value of $\Delta D_{B}$ was defined as the difference
in $D_{B}$ (Balmer jump) for a given star from that of a standard (non-accreting)
star of the same spectral type, where $\Delta D_{B}$ is the difference in the
magnitudes at both sides of the discontinuity.The relationship between $\Delta
D_{B}$ and $L_{\rm acc}$ calculated by Muzerolle et al.\ (\cite{Muzerolle2004}) and between $L_{\rm acc}$
and ${\dot{M}}_{\rm acc}$ given in Sect.~\ref{sect:calib_l} were used by
Donehew \& Brittain (\cite{DonehewBrittain2011}) to determine ${\dot{M}}_{\rm acc}$ of their program stars with
different $M_{*}$ and $R_{*}$. They used the expression:

\begin{equation}
\Delta D_{B} = 2.5 \log\frac{F_{*}+F_{A}}{F_{*}}.
\end{equation}

The observational $\Delta D_{B}$ for all targets have been measured in the observed
spectra normalized to the flux $F_{\lambda}$ at 4000\,\AA{} using the spectra of
standard stars and the Castelli-Kurucz LTE models (Castelli \& Kurucz \cite{CastelliKurucz2008}) of stellar
atmospheres with corresponding $T_{\rm eff}$ and $\log\,g$. As it turned out, the
observational $\Delta D_{B}$ demonstrated a very convincing correlation with
theoretical ${\dot{M}}_{\rm acc}$. This result is likely to be related to the fact that the $M_{*}$/$R_{*}$-ratio is practically
the same in PMS stars from late-B to K types with a dispersion of $\pm$0.1\,dex. We have built the dependence between $\Delta
D_{B}$ and ${\dot{M}}_{\rm acc}$ using the data taken from Columns 2 and 3 of
Table~3 in Donehew \& Brittain (\cite{DonehewBrittain2011}) and fitting them by a polynomial. This relation is shown
in Fig.~\ref{f1} of this paper and used in our study.

Donehew \& Brittain (\cite{DonehewBrittain2011}) measured also the equivalent widths of the Br${\gamma}$ emission
line and constructed the empiric correlation between $\log(L_{\rm acc})$ (determined
from $\Delta D_{B}$) and $L_{{\rm Br}\gamma}$. They concluded that this relationship
is different for HAes and for  earlier type HAeBes (HBes) and that the correlation
for HAes is in satisfactory agreement with the calibration constructed earlier by
Calvet et al.\ (\cite{Calvet2004}) for classical TTS (CTTS) and IMTTS. Therefore, it appears that the
Br$\gamma$ emission line is probably a reliable tracer of ${\dot{M}}_{\rm acc}$ in HAes. One of the
goals of our study is testing the reliability of additional available spectral
indicators.

Also Garcia Lopez et al.\ (\cite{GarciaLopez2006}) determined ${\dot{M}}_{\rm acc}$ for 36 HAes using a similar
calibration for TTS. Eight objects are in common between their target list and the
list of Donehew \& Brittain (\cite{DonehewBrittain2011}). For four objects the agreement in the derived
${\dot{M}}_{\rm acc}$ values is rather good with $\Delta$ $\log{\dot{M}}_{\rm acc}$
ranging from 0.01 to 0.23\,dex. Four other objects demonstrate significant
differences from 0.7 to 1.2\,dex.
 As a whole, the difference between the results
of these two works makes up $-0.18\pm0.75$\,dex, where the error is the standard
deviation. Taking into account that the accuracy of an individual estimate is of
the order of $\pm$0.4--0.5\,dex in both papers, the conclusion can be made that the results
obtained in these two works demonstrate no significant difference.


 During the time when our work was in preparation one additional paper was published by
Mendigutia et al.\ (\cite{Mendigutia2011a}), where a similar model method was used to determine mass
accretion rates in 38 HAeBes and the L-type calibrational relationships were
constructed for several emission lines. We compared the results of this work
with those obtained in Donehew \& Brittain (\cite{DonehewBrittain2011})
and Garcia Lopez et al.\ (\cite{GarciaLopez2006}). The Mendigutia et al.\ (\cite{Mendigutia2011a})
sample has overlap with the target list of Donehew \& Brittain (\cite{DonehewBrittain2011}) -- 14 stars -- and with
that of Garcia Lopez et al.\ (\cite{GarciaLopez2006}) -- 13 stars. We see that ${\dot{M}}_{\rm acc}$ of about half
the targets of Mendigutia et al.\ (\cite{Mendigutia2011a}) are estimated to be systematically higher in
comparison with those obtained in the two other studies ($+0.74\pm0.44$\,dex for
Donehew \& Brittain \cite{DonehewBrittain2011} and $+1.67\pm0.83$\,dex for Garcia Lopez et al.\ \cite{GarciaLopez2006}). Only upper limits of
${\dot{M}}_{\rm acc}$ have been obtained in Mendigutia et al.\ (\cite{Mendigutia2011a}) for the second half
of their targets. The differences in their ${\dot{M}}_{\rm acc}$ determinations with
those obtained in the two other works account for more than $-0.74\pm0.56$\,dex compared to the work of
Donehew \& Brittain (\cite{DonehewBrittain2011}) and $-0.50\pm0.24$\,dex compared to the work of Garcia Lopez et al.\ (\cite{GarciaLopez2006}).

Trying to understand the possible
cause of these discrepancies, we compared the values of the emission Balmer jump
$\Delta D_{B}$ measured in Mendigutia et al.\ (\cite{Mendigutia2011a}) by a photometric method and in
Donehew \& Brittain (\cite{DonehewBrittain2011}), where a spectroscopic method was used. We have obtained the
difference $0.17\pm0.17$. Taking into account the different dates of observations and
different methods of
measurements and their errors ($\pm0.07$ for Mendigutia et al.\ \cite{Mendigutia2011a} and $\pm0.03$ for
Donehew \& Brittain \cite{DonehewBrittain2011}), we can conclude that the differences in these measurements are not essential.

Thus we assume that some systematic distinctions can be expected in a model calculating the
relationship between ${\dot{M}}_{\rm acc}$ and $\Delta D_{B}$ carried out in the different studies.
We note that ${\dot{M}}_{\rm acc}$ values determined in
Mendigutia et al.\ (\cite{Mendigutia2011a}) appear significantly overestimated: ten objects in their study
present rates ranging from $10^{-4}$ to $10^{-6}$\,$M_{\sun}$/yr. Such large mass
accretion rates can be expected only in protostars or in such unique objects as
FUORs. In addition, according to the stellar masses and ages presented in Table~1 of
Mendigutia et al.\ (\cite{Mendigutia2011a}), six out of ten targets of the spectral type B8 -- A2  are located on the HR
diagram close to the main sequence where the accretion rates already become smaller
than at earlier stages of the PMS evolution. Generally, the masses of accretion disks around HAeBes
range from $0.01$ to $0.1M_{\sun}$
(Hillenbrand et al.\ \cite{Hillenbrand1998}, Henning et al.\ \cite{Henning1998}, Natta et al.\ \cite{Natta2000}). With mass accretion rates of the order of
$10^{-4}$ to $10^{-6}$\,$M_{\sun}$/yr, the accretion disk would  be
completely dissipated already after $10^{2}-10^{5}$\,years. This time interval is much smaller
than the time of the whole PMS evolution stage of objects with masses of $2M_{\sun}$, which is
about $10^{6}$\,years according to Palla \& Stahler (\cite{PallaStahler1993}).

 Based on these results, the question arises whether calibrations
determined for lower mass pre-main sequence stars, T\,Tauri stars, can be employed to estimate
${\dot{M}}_{\rm acc}$ in HAeBes. Clearly, applicability of such calibrations  for
${\dot{M}}_{\rm acc}$ in HAeBes demands a special examination. It can be assumed that
identical calibrations are applicable to HAeBes provided that the character of the
disk/magnetosphere and the disk/wind interaction in TTS and HAeBes is similar and
that the contribution of the CS regions to the whole CS spectrum formed in TTS and
HAeBes is comparable too. As was discussed in Sect.~\ref{sect:intro}, the size of
the magnetosphere and the value of $B$ is expected to be much smaller in HAeBes in
comparison with TTS. It is not clear yet whether these differences have an impact on the
relations between
${\dot{M}}_{\rm acc}$ and the emerging radiation in spectral lines.

\section{Observations and data analysis}
\label{sect:observations}

\begin{table}
\caption{
Target stars for which X-shooter data were obtained during our
observing run.
Spectral type and photometric data were taken from the SIMBAD database.
}
\label{Table_1} \centering
\begin{tabular} {@{}lcccccc@{}}
\hline \noalign{\smallskip}
Object&Spectral &V&R&J&H&$A_{V}$\\
&type&&&&&\\
\hline \noalign{\smallskip}
HD\,97048   & A0pshe & 8.46 & 8.50 & 7.27 & 6.67 & 1.00 \\
HD\,100546  & B9Vne  & 6.80 & 6.70 & 6.43 & 5.96 & 0.10 \\
HD\,101412  & B9.5V  & 9.29 & 9.30 & 8.64 & 8.22 & 0.54 \\
HD\,135344B & F4--F8 & 7.85 & 7.83 & 7.31 & 6.67 & 0.10 \\
HD\,150193  & A1Ve   & 8.88 & 8.90 & 6.95 & 6.21 & 1.60 \\
HD\,176386  & B9IVe  & 7.30 &  --  & 6.90 & 6.75 & 0.60 \\
HD\,190073  & A2IVpe & 7.82 & 7.80 & 7.19 & 6.65 & 0.12 \\
PDS\,2      & F2     & 10.73&  --  & 10.01& 9.68 & 0.55 \\
\noalign{\smallskip} \hline
\end{tabular}
\end{table}

\begin{table*}
\begin{flushleft}
\centering \caption{Stellar parameters of the targets.
The majority of the data are taken
from Hubrig et al.\ (\cite{Hubrig2009}). Further data is from; a -- Catala et al.\ (\cite{Catala2007});
b -- Montesinos et al.\ (\cite{Montesinos2009}); c -- Coulson \& Walter (\cite{CoulsonWalter1995}); d -- this work. }
\label{Table_2}
\begin{tabular} {lcccccc}
 \hline
 \noalign{\smallskip}
 Object & $T_{\rm eff}$ & log\,$g$ & $M_{*}/M_{\sun}$ & $R_{*}/R_{\sun}$ & $v$\,sin$\,i$ & $<B_{\rm z}>$ \\
        &       [K]            &          &                  &                &  {km/s} & [G]   \\
\hline \noalign{\smallskip} HD\,97048   & 10000   & 4.0 & 2.5    & 2.2 & 140  & 188$\pm$47 \\
HD\,100546  & 10500   & 4.5 & 2.5    & 1.5  & 65 & 89$\pm$26  \\
HD\,101412  &  9500   & 4.0 & 2.5    & 2.7  & 5 & -454$\pm$42    \\
HD\,135344B &  $6250^c$ & $4.0^c$ & 1.36 & 1.9  & 70 & -37$\pm$12   \\
HD\,150193  &  9000   & 4.0 & 2.2    & 2.0 & 100 & -252$\pm$48   \\
HD\,176386  & 10000   & 4.0 & 2.7    & 2.3 &   & -121$\pm$35 \\
HD\,190073  &  9250   & 3.5 & $2.85^a$ & $3.6^a$ & 12 & 104$\pm$19  \\
            &         &     &        & $2.1^b$ & &    \\
PDS\,2      &  $7000^d$   & $4.0^d$ & 2.5    & 1.6 & $30^d$ & 103$\pm$29  \\
\hline
\end{tabular}
\end{flushleft}
\end{table*}

\begin{table}
\label{tab:Table_3}
\caption{ Non-accreting stars used as flux standards }
\centering
\begin{tabular} {lccc}
\hline \noalign{\smallskip}
Object&Spectral type & MJD& V \\
 \hline \noalign{\smallskip}
HD\,100604  & F2V   & 55352.008 & 7.69  \\
HD\,100627  & F6IV/V & 55408.332 & 8.49 \\
HD\,100926  & A3III/IV & 55466.170 & 9.75 \\
HD\,100928  & A0IV   & 55648.147 & 9.52 \\
\noalign{\smallskip} \hline
\end{tabular}
\end{table}

The observations were performed in service mode with the X-shooter spectrograph
mounted on the 8\,m UT2 of the VLT. X-shooter allows to obtain spectral data
simultaneously over the entire wavelength range from the near-UV to the near-IR in
three arms (the UVB-arm covering the range 300--590\,nm, the VIS-arm 550--1000\,nm,
and the NIR-arm 1.0--2.5\,$\mu$m). The observations were performed with the highest
possible spectral resolution, i.e. $R$ is $\sim$9100 in the UVB-arm, $\sim$17\,400
in the VIS-arm, and $\sim$11\,000 in the NIR-arm (D'Odorico et al.\ \cite{DOdorico2006}). The data were
reduced using the X-shooter pipeline (Version 1.1.0) following the standard steps.
For more details see Modigliani et al.\ (\cite{Modigliani2010}).  Due to a very high efficiency
of the X-shooter spectrograph the signal-to-noise ratio (S/N) of 300--500
was achieved during exposure times ranging from 13--15\,s for the brightest targets
to 450\,s for the faintest target PDS\,2.

26 spectra of eight HAes with magnetic field detections were obtained during 13 nights
distributed between May and September 2010. After each science exposure, telluric standards
were automatically observed in Obseratory time at S/N$\sim$100. They are usually
main-sequence hot stars or solar analogs. Further, the data package delivered by ESO
includes a number of spectrophotometric standards. The flux standards used in our work
are presented in Table~\ref{tab:Table_3}.

\begin{figure}
\centering
\includegraphics[width=0.45\textwidth]{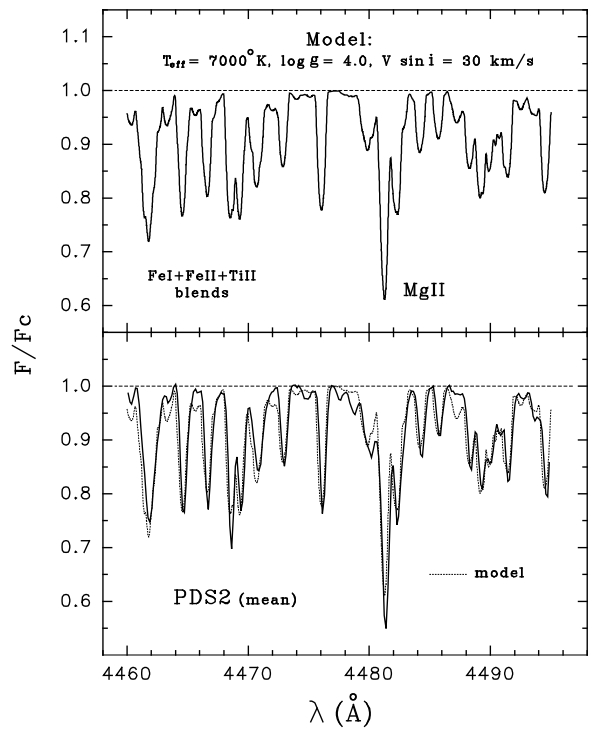}
\caption{ Comparison of the mean observed spectrum of PDS\,2 in the range of
$\lambda\lambda$4460--4500\AA{} with the synthetic spectrum corresponding to the best fit.
Top panel: The synthetic spectrum. Bottom panel: The overplotted observed spectrum (solid line) and
the synthetic spectrum (dotted line).} \label{f2}
\end{figure}

The original observation request foresaw several
observations per target over the rotation period, but due to technical problems with
X-shooter and mediocre weather conditions, the program was only partially completed.
The targets were chosen
from the sample of Herbig Ae/Be stars investigated previously by (Hubrig et al.\ \cite{Hubrig2009}).
The list of X-shooter targets is presented in Table~\ref{Table_1}, and their stellar
parameters together with the detected longitudinal magnetic fields are summarized
in Table~\ref{Table_2}. Since the star PDS\,2 was less
intensively studied in the past than other targets, we estimated atmospheric
parameters of this object by comparison between the observed spectrum in the range
$\lambda\lambda$4460--4500\,\AA{} and the  synthetic spectrum computed using the
code SYNTH+ROTATE (Piskunov \cite{Piskunov.1992}). The best model fit is presented in Fig.~\ref{f2}
with parameters: $T_{\rm eff}$= 7000\,K , log\,$g$ = 4.0, and $v$\,sin\,$i=30$\,km/s.
 The complete list of observing dates is given in Table~\ref{tab:Table_5}.

The data analysis concentrated on the emission lines and the emission Balmer jump
that can be considered as accretion indicators. To remove telluric features, we used
the telluric standards observed immediately after each observation at a similar
zenith distance. The equivalent widths (EWs) of emission lines were measured after
subtraction of stellar atmospheric profiles from the observed ones. The synthetic
atmospheric profiles were calculated using the computer code SYNTH+ROTATE
(Piskunov \cite{Piskunov.1992}) and the standard LTE Kurucz models for the corresponding values of
the stellar parameters (see Table~\ref{Table_2}). The synthetic atmospheric line
profiles were also smoothed in accordance with the spectral resolution and
$v$\,sin\,$i$. For the near-IR hydrogen lines (Br${\gamma}$, Pa$\beta$, and
Pa$\gamma$), the excess of radiation was taken into account by comparing the
photometric magnitudes of the target stars with those for unreddened stars of
corresponding spectral type and a correction for $A_{\lambda}$ using the standard
extinction law. All necessary data for this step are presented in
Table~\ref{Table_1}. The EWs of the Na~I~D lines were measured as a sum value for
both components $D_{1}$ and $D_{2}$, where the water vapour absorption lines and the
interstellar (IS) narrow components were removed prior to this step. The IS
components of the Na~I~D lines were cleaned by simple cutting. This procedure leads
to a small underestimation of the EW, but it does not account for more than 10\%
($\sim$0.1\,dex) in the value of $\log{\dot{M}}_{\rm acc}$. The emission lines of
the IR Ca~II triplet are blended by weak emission components of Pa13, Pa15, and
Pa16. But their contribution to the EWs of the Ca~II lines can be taken into account
by polynomial interpolation of the EW(Pa-lines) using synthetic spectra based on the
unblended Pa12, Pa14, and Pa17 lines (Fig.~\ref{f6}) We used standard LTE
Castelli-Kurucz models (Castelli \& Kurucz \cite{CastelliKurucz2008}) to determine fluxes at different wavelengths
corresponding to lines chosen as accretion indicators for all program stars. The
values of their stellar masses $M_{*}$ and radii $R_{*}$ were taken from
Table~\ref{Table_2} to calculate luminosities in the line-indicators.

\begin{figure}
\centering
\includegraphics[width=0.45\textwidth]{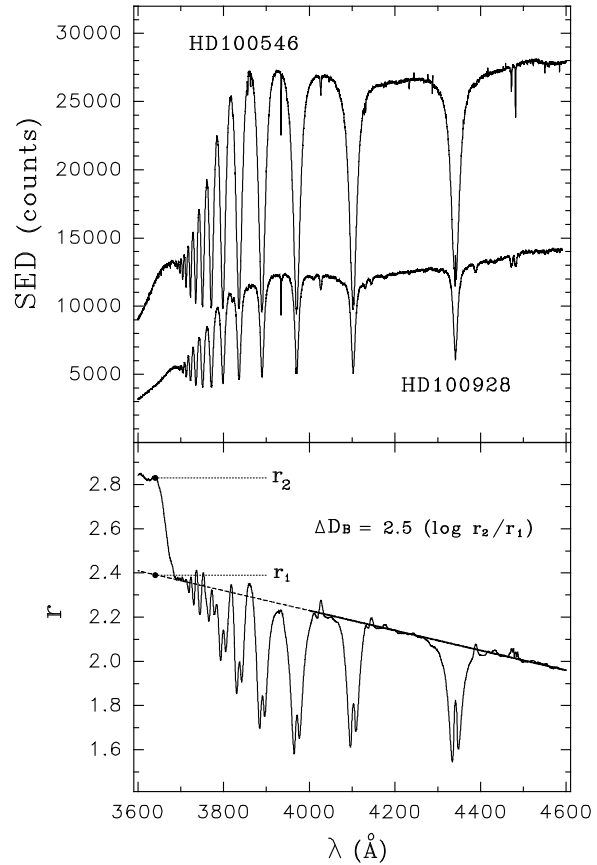}
\caption{ Spectral energy distribution (SED) in the observed spectrum of HD\,100546
and of the normal main-sequence star HD\,100928 of spectral type A0V in the range of
$\lambda\lambda$3600--4600\,\AA{} covering the Balmer jump (top panel). The bottom
panel illustrates the ratio $r$ = SED(HD\,100546)/SED(HD\,100928).
The solid line marks the
polynomial approximation of $r(\lambda$) between $\lambda$4000\,\AA{} and
$\lambda$4600\,\AA{}, and the dotted line shows its extrapolation up to
$\lambda$3640\,\AA{}. The formula for the calculation of the emission Balmer jump is also
inserted.}
\label{f3}
\end{figure}

The procedure of measuring the emission Balmer jump $\Delta D_{B}$ is illustrated in
Fig.~\ref{f3} using HD\,100546 as an example. The ratio $r$ of  the initial
unreduced spectra of the target (HD\,100546) and the standard star of spectral type
A0 (HD\,100928) is approximated polynomially between 4000\,\AA{} and 4600\,\AA{}.
The polynomial is extrapolated to 3640\,\AA{} and the ratio of ordinates $r_{2}$ and
$r_{1}$ allows to determine the value of $\Delta D_{B}$ = 2.5 log ($r_{2}$/$r_{1}$).
This procedure is identical to that used in the work of Donehew \& Brittain
(\cite{DonehewBrittain2011}).

\section{Spectral accretion indicators}
\label{sect:accr_indicators}

We examined 13 spectral diagnostics of ${\dot{M}}_{\rm acc}$ that have previously
been used in studies of TTS and BD. The basic indicator was the emission Balmer jump
$\Delta D_{B}$, calculated by the method described in Sect.~\ref{sect:observations}.
We used the relation between $\Delta D_{B}$ and ${\dot{M}}_{\rm acc}$ presented in
Fig.~\ref{f1}. Eight indicators, based on the measurements of $L_{\rm acc}$ include
H$\alpha$, H$\beta$, He~I $\lambda$5876, Br$\gamma$, Na~I~D, Pa$\beta$, Pa$\gamma$,
and Ca~II $\lambda$8542. We considered also three indicators where flux in an
emission line was used for the direct determination of ${\dot{M}}_{\rm acc}$
according to eq.\,3: He~I $\lambda$5876, Ca~II $\lambda$8542, and Ca~II
$\lambda$8662. Further in the text we use the abbreviations ``(L)'' and ``(F)'' to
mark indicators based on the measurements of $L_{\rm acc}$ and flux, respectively
(e.g., Pa$\gamma\,(L)$, Ca~II $\lambda$8662\,(F), etc.). The last indicator is
H$\alpha10\%$, which was calibrated by Natta et al.\ (\cite{Natta2004}). All these diagnostics
are presented in Table\,4.

\begin{table*}
\begin{flushleft}
\centering \caption{Mass accretion rate diagnostics that have been used in previous
studies of TTs and BDs and are analyzed in this work } \label{Table_4}
\begin{tabular} {lcc}
\hline \noalign{\smallskip}
Reference & Diagnostic & Mass range [$M_{*}/M_{\sun}$] \\
\hline \noalign{\smallskip}
Fang et al.\ (\cite{Fang2009})   & H$\alpha\,(L)$, H$\beta\,(L)$, He~I $\lambda$5876\,(L) & 0.1 -- 2.0   \\
Herczeg \& Hillenbrand (\cite{HerczegHillenbrand2008}) & Na~I~D\,(L), Ca~II $\lambda$8542\,(F), Ca~II $\lambda$8662\,(F),
He~I $\lambda$5876\,(F) & 0.1 -- 1.0\\
Dahm (\cite{Dahm.2008})  & Pa$\beta$\,(L), Pa$\gamma$\,(L), and Ca~II $\lambda$8542\,(L) & 0.5 -- 2.0 \\
Natta et al.\ (\cite{Natta2004})  & Pa$\beta$\,(L), H$\alpha10\%$ & 0.01 -- 0.1  \\
Gatti et al.\ (\cite{Gatti2008})  & Pa$\beta$\,(L), Pa$\gamma$\,(L) &  0.1 -- 1.0  \\
\hline
\end{tabular}
\end{flushleft}
\end{table*}

\begin{figure*}
\centering
\includegraphics[width=0.8\textwidth]{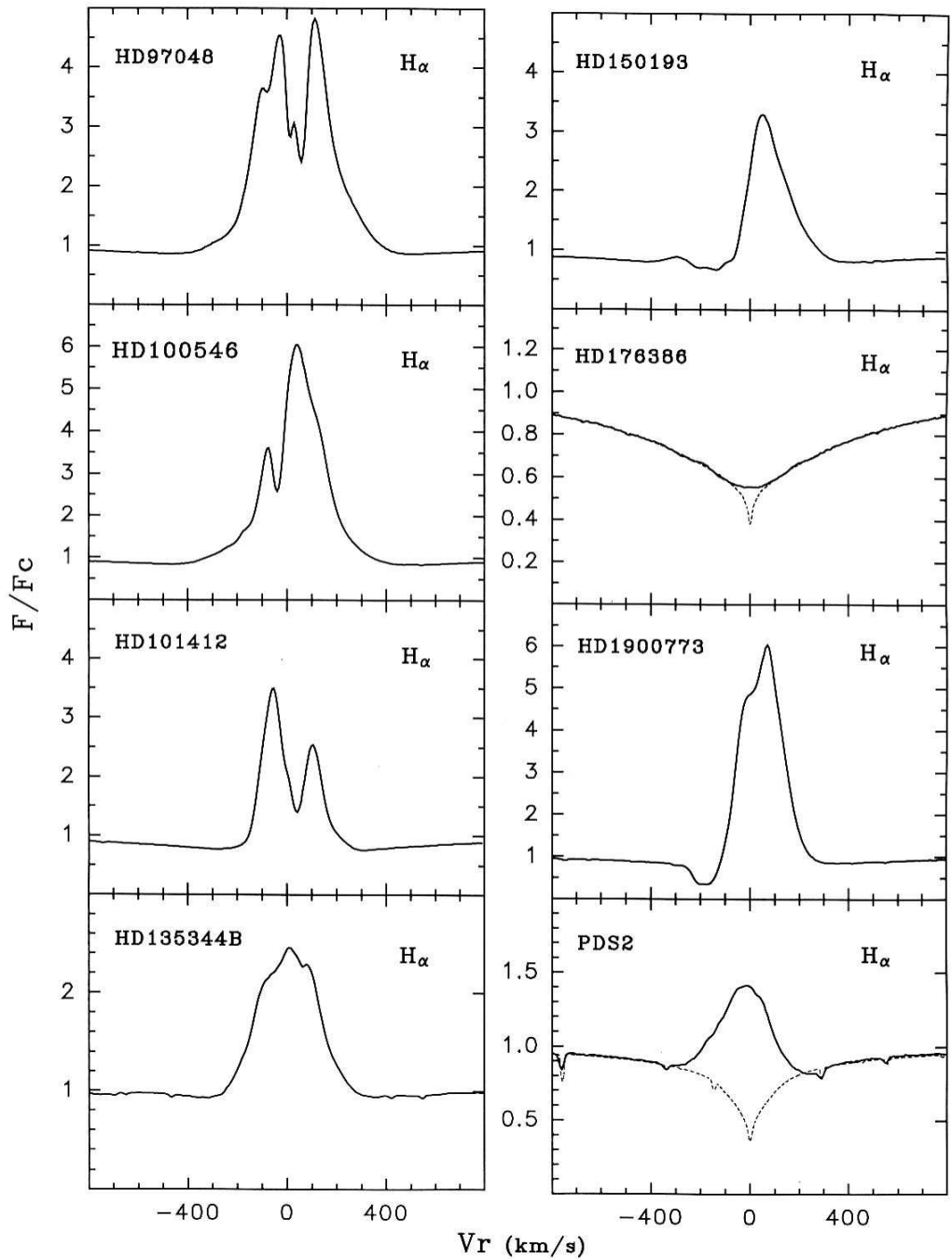}
\caption{ Normalized H$\alpha$ profiles in the spectra of studied HAes.
For HD\,135344B, HD\,150193, HD\,176386, HD\,190073, and
PDS\,2 the displayed profiles are observed on dates MJD55357.329.371,  55352.047,
55465.138, 55446.078, and 55410.343, respectively.
Profiles of HD\,135344B and PDS\,2 are shown in comparison with
synthetic profiles calculated with the computer code SYNTH+ROTATE
of (Piskunov \cite{Piskunov.1992}). The LTE models of Castelli \& Kurucz (\cite{CastelliKurucz2008}) are
used for the stellar parameters listed in Table~\ref{Table_2}. In
the case of PDS\,2, the value $v$\,sin\,$i=30$\,km/s was used to
calculate the model profile. } \label{f4}
\end{figure*}

\begin{figure*}
\centering
\includegraphics[width=0.8\textwidth]{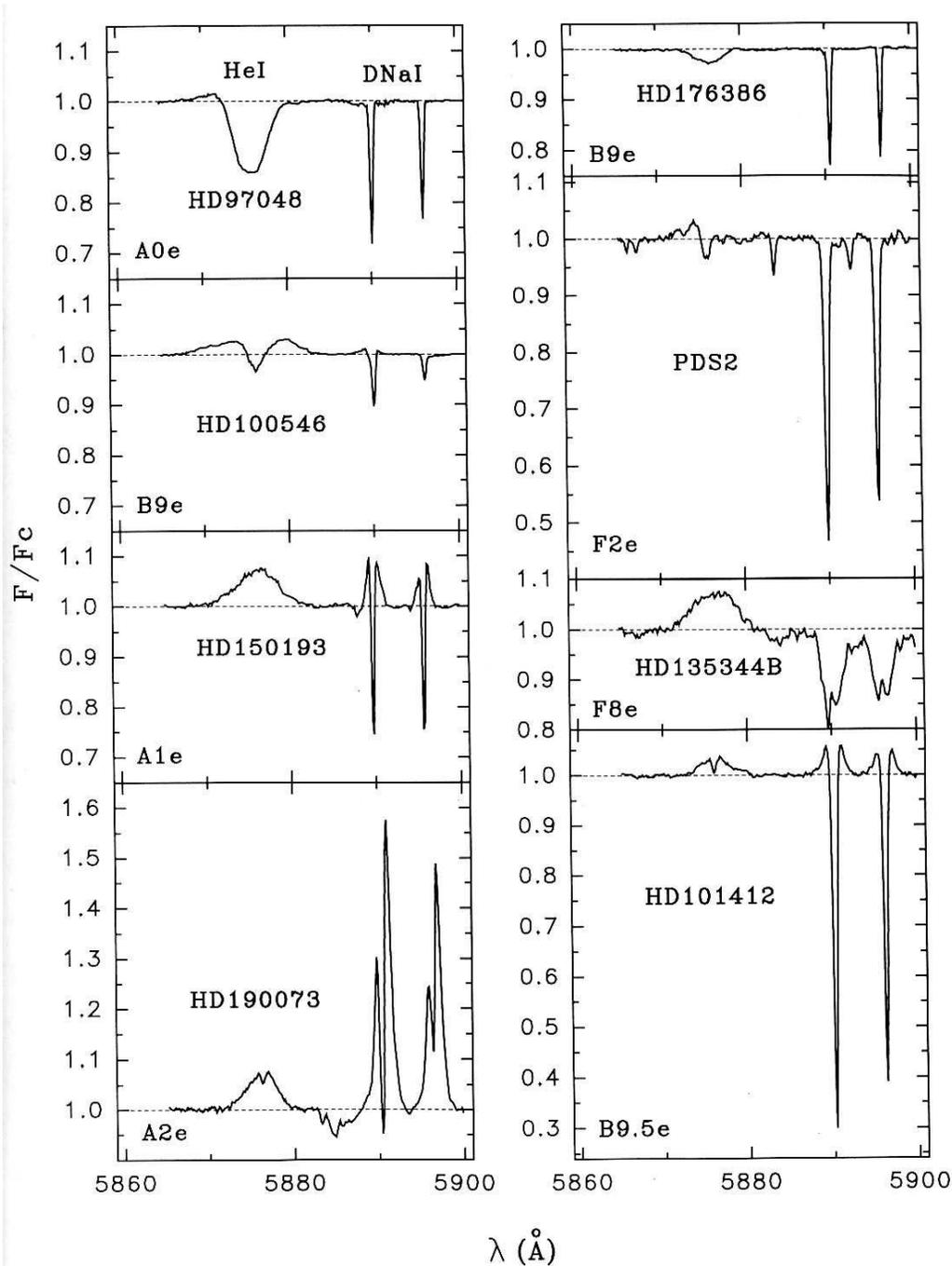}
\caption{ The same as in Fig.~\ref{f4}, but for the spectral region around He~I 5876
and Na~I~D. } \label{f5}
\end{figure*}

\begin{figure}
\centering
\includegraphics[width=0.4\textwidth]{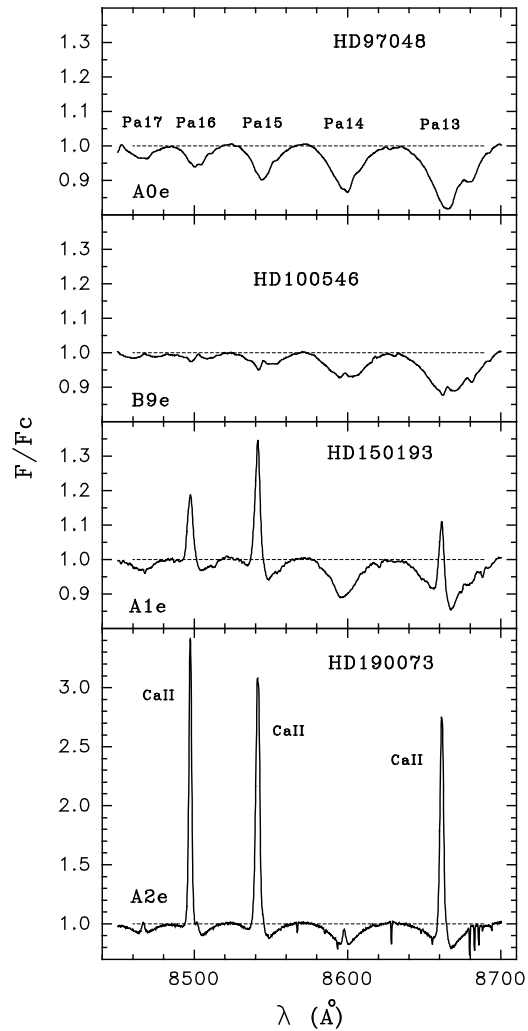}
\caption{ Normalized spectra of HD\,97048, HD\,100546, HD\,150193, and HD\,190073
in the spectral region containing near-IR Ca triple lines. The positions of the Pa lines are
indicated in the
upper panel, and the positions of the Ca~II lines in the lower panel. Strong emission near-IR Ca triplet
lines appear only in two HAe stars, HD\,150193 and HD\,190073.} \label{f6}
\end{figure}

Selected regions of the observed spectra of the program stars are presented in
Figs.~\ref{f4} to \ref{f7}.

\subsection{H$\alpha$ emission}
\label{sect:halpha}

The emission H$\alpha$ profiles vary strongly from object to object (Fig.~\ref{f4})
. HD\,150193 and HD\,190073 demonstrate P\,Cyg-type profiles indicating the
existence of a stellar wind between the star and the observer. The H$\alpha$
profiles of HD\,100546 and HD\,101412 are double-peaked with a central absorption
shifted relative to the stellar rest wavelength of the line. The H$\alpha$ profile
of HD\,97048 is double-peaked too, but of a complex structure, and single emission
profiles are observed in the spectra of HD\,135433B and PDS\,2.

As can be seen in Fig.~\ref{f4}, the H$\alpha$ emission is practically invisible in
the spectrum of the B9e star HD\,176386. This fact was already discussed by other
authors. Bibo et al.\ (\cite{Bibo1992}) characterized this object as a higher-mass analog of so-called
weak-line T\,Tauri stars (WTTS) with already dispersed accretion disks. On the other
hand, Grady et al.\ (\cite{Grady1993}) report that signatures of a matter infall onto the star is
observed in the UV spectrum of HD\,176386. If the disk of HD\,176386 is already dispersed, calibrations
derived for the magnetospheric accretion scenario are no longer valid because the balance of the emission
from different parts of the circumstellar components (magnetosphere, disk, wind) changes. In any case, the
standard model of magnetospheric accretion from the disk can hardly be applied to this object.
Therefore, any empirical spectral calibration based on the MA model cannot be used.

\subsection{Other emission lines}
\label{sect:otherlines}

Fig.~\ref{f5} illustrates the spectra of the targets in the region of the He~I
$\lambda$5876 and  Na~I~D lines, and Fig.~\ref{f6} in the region of the IR Ca~II
triplet. The emission profiles of He~I $\lambda$5876 are in general single-peak or
double-peak type in the spectra of the HAeBes, but in some cases a deep redshifted
absorption overlaps the red emission wing (see for more details
Bohm \& Catala \cite{BohmCatala1995}, Mendigutia et al.\ \cite{Mendigutia2011b}). An example of such a feature is seen in the spectrum of
HD\,97048 (Fig.~\ref{f5}). This line originates in the highest temperature regions
of the CS envelope involving the inner boundary of the accretion disk interacting
with the magnetosphere, the matter streams infalling onto the star, and the region
of the impact of the streams on the stellar surface. These regions are rather
compact in size and screening of the stellar limb by the accreted flows can be an
important factor in forming the observed line profile. This screening leads to the
appearance of a redshifted absorption component of the profile. Therefore, we have
to be careful in using the He~I emission for the ${\dot{M}}_{\rm acc}$
determination. At the time when the infalling stream screens the stellar limb, the
${\dot{M}}_{\rm acc}$ measurement can be considerably underestimated.

The types of the Na~I~D line profiles (Fig.~\ref{f5}) as well as of the IR Ca~II
triplet profiles (Fig.~\ref{f6}) in the spectra of the targets are very diverse.
Even in objects of similar spectral types (A2--B9) are the intensities of these
emission lines completely different.
The emission in the Na~I~D lines is clearly visible and the EW can be measured in
three objects (HD\,190073, HD\,150193, and HD\,101412). As for the Ca~II triplet,
the EW can be measured only in HD\,190073 and HD\,150193. It is remarkable that in
our sample HD\,190073 is the star with the strongest intensities of the Na~I~D and
IR Ca~II lines. This object demonstrates the clearest developed P\,Cyg-structure of
the H$\alpha$ profile (Fig.~\ref{f4}). The second object with an H$\alpha$ profile
of P\,Cyg-type is HD\,150193. It shows also rather intense emission in Na~I~D and IR
Ca~II. This fact leads us to the hypothesis that the appearance of considerable
emission in the Na~I~D and IR Ca~II lines is related to the presence of a wind
zone between the star and the observer. The signatures of the matter outflow are
clearly seen in the Na~I~D lines of HD\,190073 and HD\,150193 in form of the
P\,Cyg-structure.

\begin{figure*}
\centering
\includegraphics[width=0.8\textwidth]{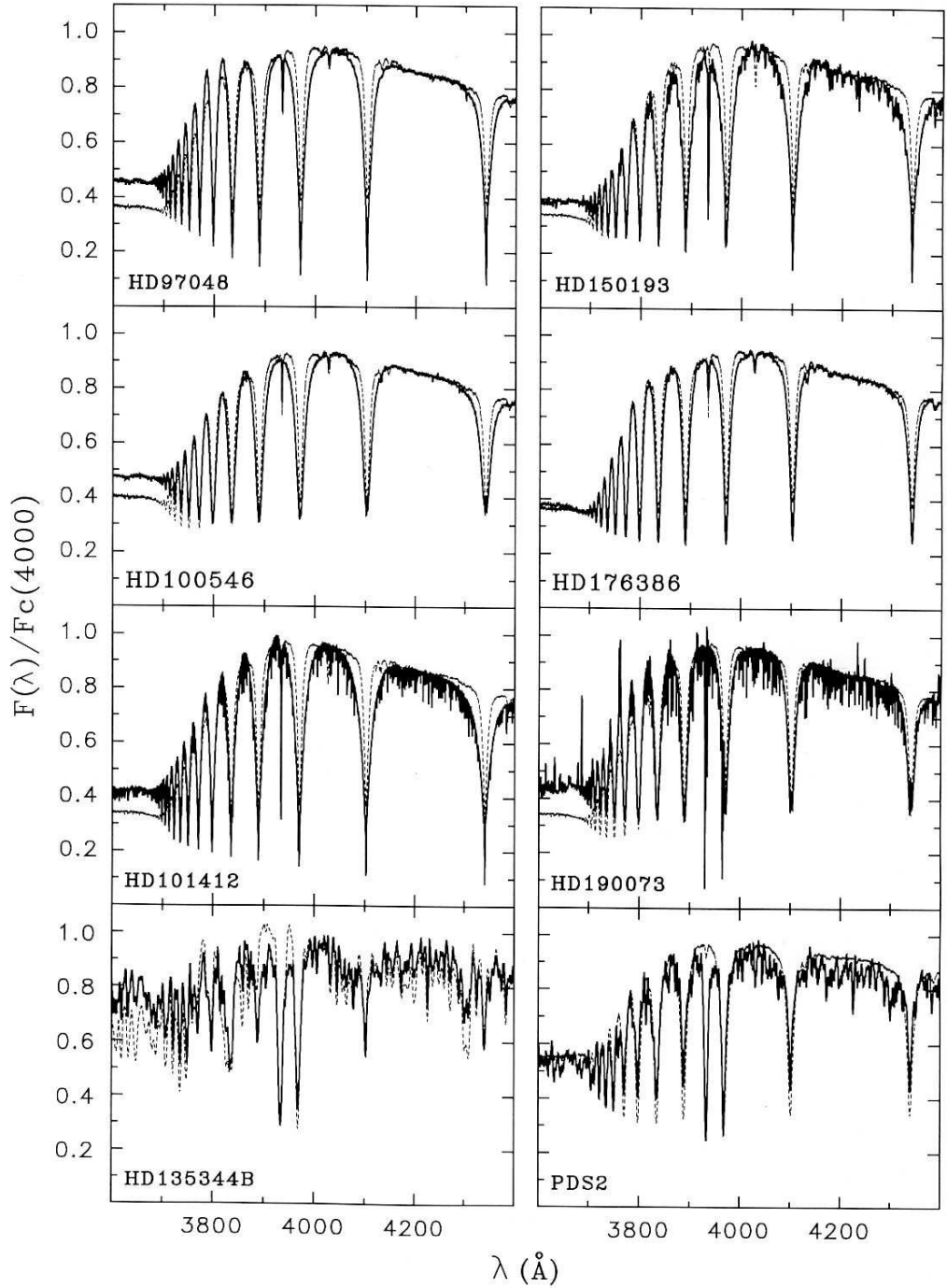}
\caption{ Normalized spectra of our targets in the region near 3600--4400\,\AA{}.
The normalization was made to the flux $F(\lambda)$ at the level of the model
continuum at $\lambda=4000$\,\AA{}. Spectra of standards of similar spectral type
are given for comparison (dashed lines). } \label{f7}
\end{figure*}

\begin{table*}
\begin{flushleft}
\centering \caption{Observing dates, magnetic/rotation phases calculated according
to the ephemerides presented in Hubrig et al.\ (\cite{Hubrig2011a}, \cite{Hubrig2011b}), and measurement
results of EWs, $W({\rm H}\alpha10\%)$, and $\Delta D_{B}$ for HAes in our sample.}
\label{tab:Table_5}
\begin{tabular}{lccccccccccccc}
 \hline
 \noalign{\smallskip}
Object & MJD & Magn. &               &              &               &      & EW [\AA{}]&              &      &       &             & $W({\rm H}\alpha10\%)$ & $\Delta D_{B}$  \\
       &     & phase & Br$\gamma$ & Pa$\beta$ & Pa$\gamma$ & Ca~II & Ca~II       & H$\alpha$ & Na~I~D & He~I   & H$\beta$ &        [km/s]               &   \\
       &     &       &               &              &               & 8662 & 8542       &              &      & 5876  &             &                             &   \\
\hline
 \noalign{\smallskip}
HD\,97048   & 55352.000 & 0.43 & 10.30 & 21.90 & 9.12  & --   & --   & 30.8  & --    & 0.0525 & 1.58 & 604 & 0.25  \\
HD\,100546  & 55352.022 &      & 16.05 & 26.41 & 9.72  & --   & --   & 30.7  & 0.035 & 0.162  & 2.55 & 540 & 0.18  \\
HD\,101412  & 55351.992 & 0.71 &  5.79 &  7.77 & 3.63  & --   & --   & 13.6  & 0.340 & 0.145  & 1.69 & 560 & 0.225 \\
HD\,135344B & 55329.362 &      &  2.27 &  3.88 & 1.47  & --   & --   & 10.3  & --    & 0.43   & 1.57 & 496 & 0.15  \\
            & 55352.034 &      &  3.21 &  4.10 & 1.72  & --   & --   & 14.5  & --    & 0.32   & 2.16 & 501 & 0.15  \\
            & 55357.251 &      &  2.65 &  3.21 & 1.39  & --   & --   & 11.1  & --    & 0.26   & 2.00 & 501 & 0.15  \\
HD\,150193  & 55329.371 & 0.59 &  8.83 &  7.66 & 4.93  & 0.87 & 1.61 & 11.9  & 0.253 & 0.530  & 1.15 & 517 & 0.19  \\
            & 55352.047 & 0.81 & 10.19 &  9.80 & 6.06  & 1.06 & 2.13 & 13.6  & 0.479 & 0.479  & 1.50 & 561 & 0.19  \\
HD\,176386  & 55329.387 & 0.09 & 0.392 & 0.188 & --    & --   & --   & 0.167 & --    & --     & --   & --  & --    \\
            & 55463.171 & 0.88 & 0.184 & 0.144 & --    & --   & --   & 0.159 & --    &   --   & --   & --  & --    \\
            & 55465.138 & 0.06 & 0.270 & 0.203 & --    & --   & --   & 0.175 & --    & --     & --   & --  & --    \\
            & 55466.166 & 0.21 & 0.231 & 0.153 & --    & --   & --   & 0.161 & --    & --     & --   & --  & --    \\
HD\,190073  & 55329.397 &      &  9.97 & 15.8  & 8.04  & 5.50 & 6.58 & 23.9  & 1.94  & 0.213  & 2.64 & --  & 0.19  \\
            & 55410.250 &      & 10.67 & 18.8  & 9.45  & 5.91 & 7.29 & 26.8  & 2.25  & 0.337  & 3.54 & --  & 0.27  \\
            & 55446.078 &      & 10.97 & 19.1  & 9.74  & 5.48 & 6.87 & 25.2  & 2.14  & 0.471  & 3.58 & --  & 0.27  \\
            & 55463.182 &      &  8.91 & 18.1  & 8.36  & 4.65 & 5.93 & 20.5  & 1.83  & 0.364  & 2.37 & --  & 0.185 \\
            & 55465.159 &      &  8.65 & 17.2  & 8.67  & 4.73 & 5.92 & 20.5  & 1.82  & 0.257  & 2.56 & --  & 0.19  \\
            & 55468.141 &      &  8.40 & 16.4  & 7.82  & 4.53 & 5.56 & 20.5  & 1.79  & 0.317  & 2.18 & --  & 0.19  \\
PDS\,2      & 55375.418 &      &  1.60 & 0.627 & 0.358 & --   & --   & 3.16  & --    & 0.0286 & 0.34 & 380 & --    \\
            & 55408.317 &      & --    & --    & --    & --   & --   & 5.69  & --    & 0.0745 & 0.68 & 348 & --    \\
            & 55409.410 &      & --    &  --   & --    & --   & --   & 5.52  & --    & 0.0585 & 0.54 & 380 & --    \\
            & 55410.343 &      & --    & --    & --    & --   & --   & 4.50  & --    & 0.0464 & 0.32 & 396 & --    \\
            & 55411.394 &      &  2.85 & 2.05  & 0.829 & --   & --   & 5.90  & --    & 0.1068 & 0.74 & 365 & --    \\
            & 55446.307 &      &  2.33 & 1.00  & 0.450 & --   & --   & 3.30  & --    & 0.0621 & 0.28 & 424 & --    \\
            & 55463.210 &      &  2.77 & 1.30  & 0.627 & --   & --   & 5.25  & --    & 0.0734 & 0.55 & 425 & --    \\
            & 55464.208 &      &  1.79 & 0.95  & 0.460 & --   & --   & 6.10  & --    & 0.0335 & 0.33 & 409 & --    \\
\noalign{\smallskip} \hline
\end{tabular}
\end{flushleft}
\end{table*}

\begin{table*}
\begin{flushleft}
\centering \caption{Mean errors of measurements presented in Table~\ref{tab:Table_5}.}
\label{Table_6}
\begin{tabular}{lccccccccccc}
 \hline
\noalign{\smallskip}
Object  &               &              &               &      &         $\pm\sigma$(EW) [\AA{}]&              &      &       &             & $W({\rm H}\alpha10\%)$ & $\Delta D_{B}$  \\
        & Br$\gamma$ & Pa$\beta$ & Pa$\gamma$ & Ca~II & Ca~II       & H$\alpha$ & Na~I~D & He~I   & H$\beta$ &        [km/s]               &  [mag] \\
        &               &              &               & 8662 & 8542       &              &      & 5876  &             &                             &   \\
\hline
\noalign{\smallskip}
HD\,97048   & 0.25 & 0.40 & 0.25  & --   & --   & 0.45  & --    & 0.015 & 0.10 & 20 & 0.03  \\
HD\,100546  & 0.30 & 0.40 & 0.25  & --   & --   & 0.45  & 0.015 & 0.020 & 0.15 & 20 & 0.03  \\
HD\,101412  & 0.20 & 0.25 & 0.20  & --   & --   & 0.30  & 0.05  & 0.030 & 0.10 & 20 & 0.03 \\
HD\,135344B & 0.15 & 0.15 & 0.10  & --   & --   & 0.25  & --    & 0.045 & 0.10 & 20 & 0.03  \\
HD\,150193  & 0.25 & 0.25 & 0.20  & 0.07 & 0.10 & 0.25  & 0.04  & 0.060 & 0.10 & 20 & 0.03  \\
HD\,176386  & 0.08 & 0.05 & --    & --   & --   & 0.05 & --     & --    & --   & --  & --    \\
HD\,190073  & 0.25 & 0.30 & 0.25  & 0.15 & 0.20 & 0.40  & 0.10  & 0.04  & 0.15 & --  & 0.03  \\
PDS\,2      & 0.10 & 0.10 & 0.06  & --   & --   & 0.20  & --    & 0.02  & 0.06 & 20 & --    \\
 \noalign{\smallskip} \hline
\end{tabular}
\end{flushleft}
\end{table*}

\subsection{Balmer jump}
\label{sect:jump}

Fig.~\ref{f7} illustrates the normalized spectra of the targets in the region of the
Balmer jump in comparison with spectra of non-accreting stars of similar spectral
types.  Castelli-Kurucz LTE (Castelli \& Kurucz \cite{CastelliKurucz2008}) models were applied for normalization
of the observed spectra of non-accreting stars to flux $F_{\lambda}$ at 4000\,\AA{}
at the continuum level. The $r$ - dependencies (see Sect.~\ref{sect:observations}
and Fig.~\ref{f3}) were used to construct the normalized spectra of the targets.

The EWs of all studied lines, the H$\alpha10\%$ width and $\Delta D_{B}$ are listed
in Table~\ref{tab:Table_5} together with the observing dates and magnetic/rotation
phases calculated according to the ephemerides presented in Hubrig et al.\
(\cite{Hubrig2011a}, \cite{Hubrig2011b}).  Mean accuracies of the measurements are
presented in Table~\ref{Table_6}. The errors of the individual estimates of the
equivalent widths
   were computed using a program that takes into account the
   width and the maximum intensity of an emission line as well as the
   S/N ratio of the spectrum near the measured line. Further, the program considers
   the contribution of the
   photospheric background and the telluric line subtraction.
   Since we used for the determination of  $\Delta D_{B}$ a method
   similar to that used by Donehew \& Brittain (\cite{DonehewBrittain2011}), our error estimates
are of the same order as presented by these authors, about 0.03\,mag.
   The errors for the individual
   measurements differ from the mean values presented in Table~\ref{Table_6} by not more
   than 20\%.

\section{Accretion rates}
\label{sect:rates}

\subsection{Testing existing empirical calibrations}
\label{sect:test_empirical}

We computed the ${\dot{M}}_{\rm acc}$ values for HAes in our sample using
the measured quantities from
Table~\ref{Table_4} for indicators specified at the beginning of
Sect.~\ref{sect:accr_indicators}.

HD\,176386 was excluded from our analysis for the reason discussed
in Sect.~\ref{sect:halpha}. The formal calculation of
${\dot{M}}_{\rm acc}$ of this object leads to $\log{\dot{M}}_{\rm
acc}=-8.70$ for Br$\gamma$, $-9.03$ for Br$\beta$, and $-9.63$ for
H$\alpha$. Other indicators are absent in all observed spectra of
the star or too weak to be measured precisely.

The indicator $\Delta D_{B}$ was used only for six objects with ${\dot{M}}_{\rm
acc}>10^{-8} M_{\sun}$/yr . As was concluded in Donehew \& Brittain (\cite{DonehewBrittain2011}), this diagnostics is
inefficient for $\Delta D_{B} < 0.1^{m}$. The width $\Delta V ({\rm H}\alpha 10\%)$
cannot be measured correctly for HD\,190073 because a strong P Cyg-type structure
overlaps the blue wing of the emission H$\alpha$ profile. Taking into account that
such features are not  common to all HAes, there is a doubt if it is
reasonable to use them at all. Still, it is of interest to compare the
${\dot{M}}_{\rm acc}$ estimations derived from these diagnostics with values
obtained from other indicators.


\begin{table*}
\centering \caption{ List of diagnostics satisfying our criterion of applicability
(besides of $\Delta D_{B}$). The error of the mean value $\sigma_{m}$ indicates the
accuracy of determination.} \label{Table_7}
\begin{tabular} {llcc}
\hline \noalign{\smallskip}
Diagnostic            & Reference       & Mean value of              & Number of  \\
``$X$''     &     &``$\Delta D_{B} - X$''      & objects  \\  \\
  &                 &[dex]                       & $n$  \\
  \hline \noalign{\smallskip}
Br$\gamma$(L)          & Donehew \& Brittain (\cite{DonehewBrittain2011})  & $+0.19\pm0.15$            & 6  \\
Pa$\beta$(L)            & Dahm (\cite{Dahm.2008})    & $+0.30\pm0.09$             & 6  \\
Pa$\gamma$(L)         & Dahm (\cite{Dahm.2008})    & $+0.11\pm0.15$             & 6  \\
H$\alpha$(L)          & Dahm (\cite{Dahm.2008})    & $-0.02\pm0.12$             & 6  \\
H$\beta$(L)           & Dahm (\cite{Dahm.2008})    & $-0.13\pm0.09$             & 6  \\
He~I $\lambda$5876(L) & Fang et al.\ (\cite{Fang2009})    & $+0.17\pm0.17$             & 6 \\
Na~I~D(L)              & Herczeg \& Hillenbrand (\cite{HerczegHillenbrand2008}) & $+0.10\pm0.16$             & 3 \\
Ca~II $\lambda$8542(L)& Dahm (\cite{Dahm.2008})    & $-0.22\pm0.07$             &  2 \\
H$\alpha 10\%$        & White \& Basri (\cite{WhiteBasri2003})   & $+0.08\pm0.10$             & 5 \\
\noalign{\smallskip} \hline
\end{tabular}
\end{table*}

\subsubsection{Criterion of applicability} \label{sect:criterion}

 Our aim was to choose diagnostics that lead to results consistent with each
other, where the  basic indicator $\Delta D_{B}$ is the only tracer connected with
${\dot{M}}_{\rm acc}$ directly on the basis of model calculations.
 We introduce the criterion of applicability of a diagnostic ``$X$'' for
measuring the ${\dot{M}}_{\rm acc}$ of HAes as follows: the indicator is applicable
if the mean value of $<\log{\dot{M}_{\rm acc}}(\Delta D_{B}) - \log{\dot{M}_{\rm
acc}}(X)>$ (further in the text as ``$\Delta D_{B} - X$'') is consistent with zero
within the 1 $\sigma_{m}$ errors, calculated for our sample. Here
$\sigma_{m}$=$\sigma$/$\sqrt{n-1}$ is the standard error of the mean value and $n$
is the number of objects.

 As a result, we identified a number of diagnostics
satisfying this criterion (see Table~\ref{Table_7}). A small disagreement is present
in the Pa$\beta$ (Dahm \cite{Dahm.2008}) indicator: $+0.30\pm0.09$\,dex, and in the Ca~II
$\lambda$8542(L) (Dahm \cite{Dahm.2008}) indicator. In any case, the mean spread of all these
differences type ``$\Delta D_{B}$ -- X'' is of the order of $\pm$0.10--0.15\,dex. It
is remarkable that if the lines Na~I~D and IR Ca~II $\lambda$8542 are clearly
visible in the spectrum of an object, the ${\dot{M}}_{\rm acc}$ estimates derived
from these indicators are in a satisfactory agreement with values obtained from
other spectral diagnostics.

The three remaining indicators of ${\dot{M}}_{\rm acc}$ that make use of F-type
calibrations (Ca~II $\lambda$8542, Ca~II $\lambda$8662, and He~I
$\lambda$5876) were additionally examined because the ``$\Delta D_{B} - X$'' values
of two Ca~II (F) indicators demonstrate differences at larger amplitudes than all
other tracers presented in Table~\ref{Table_7}: ($+0.44\pm0.32$\,dex and
$-0.28\pm0.35$\,dex for Ca~II $\lambda$8542 (F) and Ca~II $\lambda$8662 (F),
respectively). Further, the He~I $\lambda$5876 tracer shows  a notable systematic shift of
``$\Delta D_{B} - X$''$ = +0.69\pm0.19$\,dex.

\subsubsection{Testing F-calibrations} \label{sect:testf}

\begin{figure*}
\centering
\includegraphics[width=0.8\textwidth]{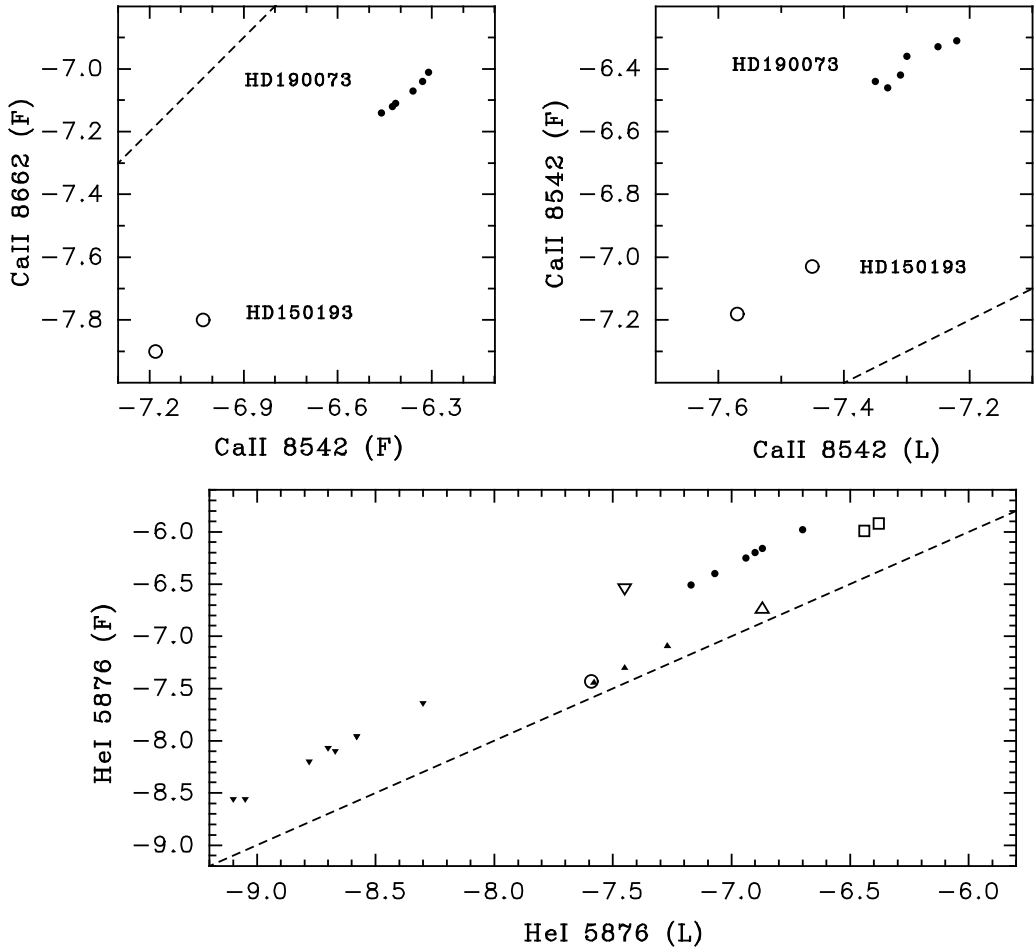}
\caption{ Illustration of systematic discrepancies between the results of the
$\log{\dot{M}}_{\rm acc}$ [dex] determination obtained from comparing three pairs of diagnostics:
Ca~II $\lambda$8542~(F) vs Ca~II $\lambda$8662~(F) (left top panel); Ca~II
$\lambda$8542~(L) vs Ca~II $\lambda$8542~(F) (right top panel); and  He~I
$\lambda$5876)~(L) vs  He~I $\lambda$5876)~(F) (bottom panel). The dashed lines
indicate the position of equality of abscissa and ordinate. Results for different
objects are marked by different symbols. In the upper panels the filled circles
mark the six measurements for HD\,190073, while open circles refer to measurements obtained
for HD\,150193. Ca line diagnostics can be used only for these two stars.
In the bottom panel HD\,97048 is marked by the open
circle, HD\,100546 by the open triangle pointing downwards, HD\,101412 by the open triangle pointing upwards, HD\,135344B by
filled triangles pointing upwards, HD\,150193 by open squares, HD\,190073 by filled circles, and PDS2 by
filled triangles pointing downwards.} \label{f8}
\end{figure*}

The analysis of ${\dot{M}}_{\rm acc}$ estimates that were obtained using these
diagnostics demonstrate considerable systematic inconsistencies between each other
as well as between F- and L-type calibrations for the same lines. These
discrepancies are illustrated in Fig.~\ref{f8}.

The estimates derived from Ca~II $\lambda$8542(F) are systematically
0.75$\pm$0.03\,dex higher than those from Ca~II $\lambda$8662(F) (top left panel).
These two lines originate in the same circumstellar region, and we conclude that
conditions in this region in HAes are different from those in TTS for which these
two F-calibrations have been introduced.

\begin{figure}
\centering
\includegraphics[width=0.4\textwidth]{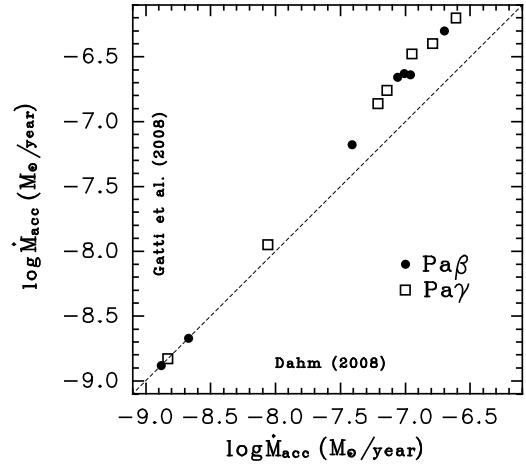}
\caption{Comparison of the results  of the $\log{\dot{M}}_{\rm acc}$ [dex]
determination for our targets obtained from the different  Pa$\beta$ and Pa$\gamma$
calibrations defined in Dahm (\cite{Dahm.2008}) and Fang et al.\ (\cite{Fang2009}). The dashed line indicates
the equality of measured values.} \label{f9}
\end{figure}

The mean difference between values obtained from $\lambda$8542(F) and
$\lambda$8542(L) is $+$0.90\,dex for HD\,190073 and $+$0.40\,dex for HD\,150193 (top
right panel). A similar picture is observed if we compare the ${\dot{M}}_{\rm
acc}$ derived with He~I $\lambda$5876(F) and He~I $\lambda$5876(L) calibrations
(bottom panel). The values obtained with the F-type calibrations are systematically
higher with the spread lying between 0.15 and 0.70\,dex for different objects. Each
star has its own offset which is the same at all observed epochs pointing at some
systematic factor. Based upon the results of this test, we assume that
the L-type calibrations are best suitable for the HAes in our sample. We suggest that the
observed systematic differences between the values derived with the F- and L-type
calibrations are likely to be caused by an insufficient accuracy of the $c$ and $d$
coefficients in the F-type calibrations applied to HAes, as well as by uncertainties
in $M_{*}$ or $R_{*}$ of the targets, used for the calculation of $L_{\rm acc}$.

\subsubsection{Other factors leading to systematic effects} \label{sect:otherf}

The uncertainties in stellar parameters (especially in the stellar radius $R_{*}$)
can be a source of considerable systematic errors in the ${\dot{M}}_{\rm acc}$
determination. As an example, we computed the mass accretion rate of HD\,190073
separately with two different values of its stellar radius: $R_{*}/R_{\sun}=3.6$
(Catala et al.\ \cite{Catala2007}) and the more recent estimate $R_{*}/R_{\sun}=2.1$ (Montesinos et al.\ \cite{Montesinos2009}).
In the first case, we have obtained a mean value of the accretion rate $-6.23$, with
a spread at the level of standard deviation of $\pm$0.17\,dex using eight spectral
indicators derived from luminosities of the emission lines (L-type). This value
strongly deviates from the $-$7.14 obtained from $\Delta D_{B}$ which, according to
Fig.~\ref{f1}, is independent of $M_{*}$ and $R_{*}$ of a star. Using
$R_{*}=2.1\,R_{\sun}$ leads to the mean value $-$7.15$\pm$0.17, that is in good
agreement with the estimation from $\Delta D_{B}$.

Applying different empiric relations between Pa$\beta$ and Pa$\gamma$ luminosity,
respectively, and the mass accretion rates that have been derived for the TTS by
various authors to our sample, we find considerable and systematic discrepancies.
This is demonstrated in Fig.~\ref{f9} where we show the ${\dot{M}}_{\rm acc}$ values
for the targets that result from the calibrations by Gatti et al.\ (\cite{Gatti2008}) versus those from
Dahm (\cite{Dahm.2008}). As can be seen in Fig.~\ref{f9}, the difference between the two
calibrations is practically absent for small $\log{\dot{M}}_{\rm acc}$ (less than
$-8.0$) and becomes up to 0.40\,dex for $\log{\dot{M}}_{\rm acc}>-7$. We have to
mention that Gatti et al.\ (\cite{Gatti2008}) have calibrated their relation over the
$\log{\dot{M}}_{\rm acc}$ range from $-9.6$ to $-8.2$ ($M_{*}$ range
0.1--1.0\,$M_{\sun}$) and Dahm (\cite{Dahm.2008}) from $-8.7$ to $-7.2$ ($M_{*}$ range
0.5--2.0\,$M_{\sun}$), that is more similar to values ${\dot{M}}_{\rm acc}$ and
$M_{*}$ for our targets. Therefore, we used in our study the diagnostics
Pa$\beta$(L) and Pa$\gamma$(L) introduced by Dahm (\cite{Dahm.2008}).

\subsection{The deduced accretion rates for the targets in our sample}
\label{sect:deduced_rates}

\begin{figure*}
\centering
\includegraphics[width=0.95\textwidth]{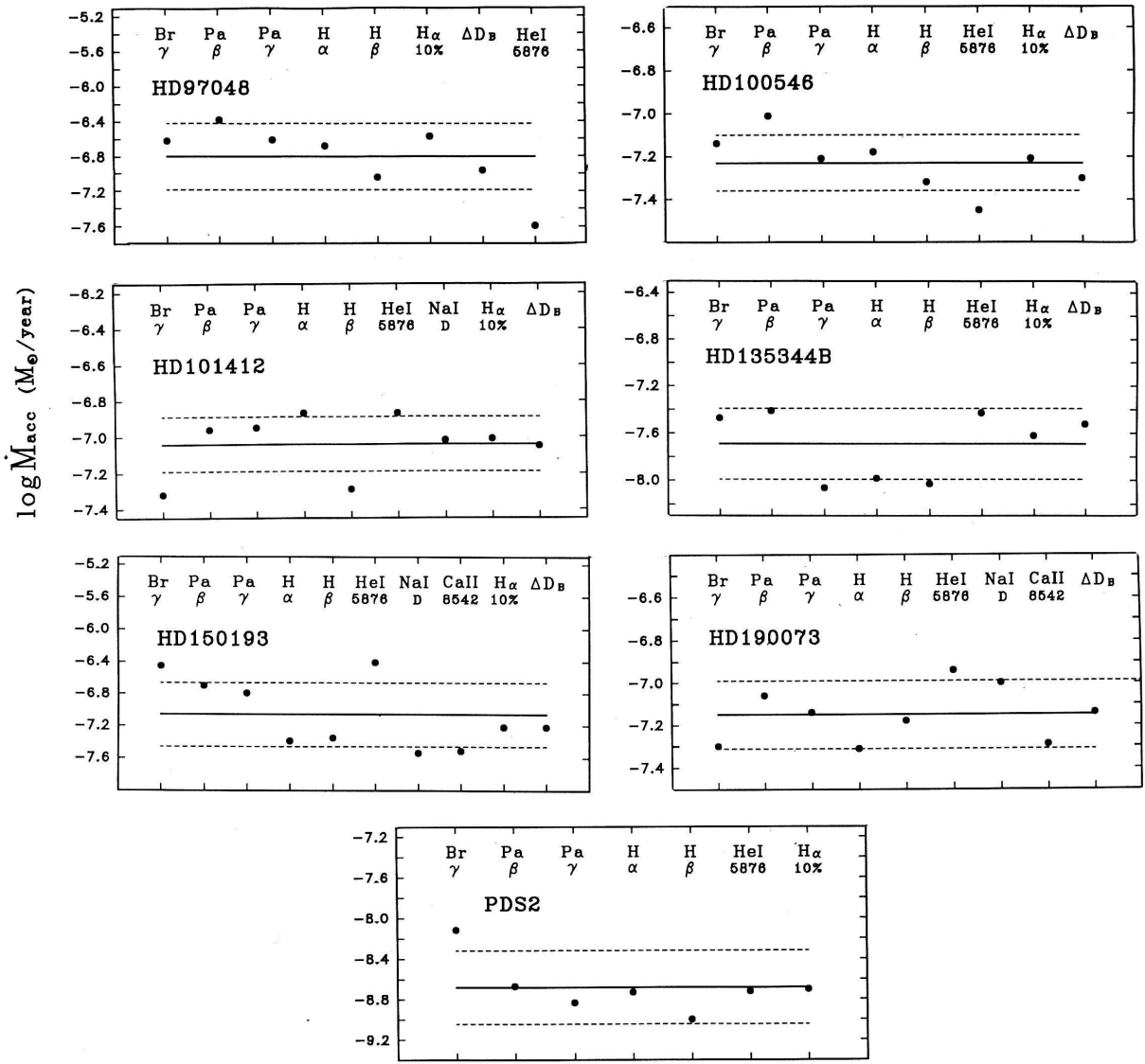}
\caption{ Mass accretion rates of the targets determined from different indicators,
as labeled. The solid line shows the mean $\log{\dot{M}}_{\rm acc}$ value and the
dashed lines indicate the region of $\pm$1$\sigma$ uncertainty where $\sigma$ is the
standard deviation.}
\label{f10}
\end{figure*}

Fig.~\ref{f10} illustrates the mass accretion rates obtained for seven program stars
(except HD\,176386) that have been derived using the indicators that passed
successfully our applicability test (see Sect.~\ref{sect:criterion}). Because the
F-type diagnostics lead to systematic differences in the ${\dot{M}}_{\rm acc}$
estimates in comparison with the values obtained with the other indicators (see
Sect.~\ref{sect:testf} and Fig.~\ref{f8}), we did not include the values of
${\dot{M}}_{\rm acc}$ derived from F-type indicators in the calculations of the mean
values and the uncertainties of the mass accretion rates. As a result, only L-type
indicators were used (except H$\alpha 10\%$ and $\Delta D_{B}$) and, further in the
text, we dropped the abbreviation (L) in the names of the indicators.

\begin{table*}
\begin{flushleft}
\centering \caption{ Mean mass accretion rates. The error is given as the standard
deviation $\sigma$. Literature data on accretion rates are taken from Garcia Lopez et al.\ (\cite{GarciaLopez2006})
and Mendigutia et al.\ (\cite{Mendigutia2011a}).} \label{Table_8}
\begin{tabular}{lccccc}
 \hline
\noalign{\smallskip}
Object & Number of   & Number of   & $\log{\dot{M}}_{\rm acc}$    & $\log{\dot{M}}_{\rm acc}$    & $\log{\dot{M}}_{\rm acc}$   \\
       & indicators  & observations   &  [$M_{\sun}/{\rm yr}$] &  [$M_{\sun}/{\rm yr}$] &  [$M_{\sun}/{\rm yr}$] \\
       &                &   &                 &  Garcia Lopez et al.\ (\cite{GarciaLopez2006}) &  Mendigutia et al.\ (\cite{Mendigutia2011a})     \\
\noalign{\smallskip} \hline \noalign{\smallskip}
HD\,97048   & 8 & 1  & $-6.80\pm0.38$  & $-7.17$  & \\
HD\,100546  & 8 & 1  & $-7.23\pm0.13$  &        & \\
HD\,101412  & 9 & 1  & $-7.04\pm0.15$  &       & \\
HD\,135344B & 8 & 3  & $-7.69\pm0.30$  & $-8.27$ &   \\
HD\,150193  & 10 & 2& $-7.06\pm0.20$  & $-7.29$ & -6.12 \\
HD\,190073  & 9 & 6  & $-7.15\pm0.16$  &      & -5.00  \\
PDS\,2      & 7 & 8   & $-8.68\pm0.36$  &     &   \\
\noalign{\smallskip} \hline
\end{tabular}
\end{flushleft}
\end{table*}

\begin{table*}
\begin{flushleft}
\centering \caption{ Mass accretion rate determinations using several diagnostics
presented in this work and the results published in the literature. }
\label{Table_9}
\begin{tabular} {ccccc}
 \hline
 \noalign{\smallskip}
&  & &Mean standard & Error of the   \\
Paper & Range of $\log{\dot{M}}_{\rm acc}$  & Range of individual  &deviation $\sigma$ &mean value  $\sigma_{m}$    \\
&       [$M_{\sun}/{\rm yr}$]             & deviations [dex]     &   [dex]     & [dex]               \\
\hline
 \noalign{\smallskip}
This paper &$-$8.68 $\div$ $-$6.69 & 0.13 -- 0.36 & 0.25 & 0.11 \\
 Dahm (\cite{Dahm.2008}) & $-$8.72  $\div$ $-$7.20 & 0.1 -- 1.4 & 0.30 & 0.12 \\
 Rigliaco et al.\ (\cite{Rigliaco2011}) & $-$9.86 & 0.4 -- 1.4 & 0.45 & 0.14 \\
\noalign{\smallskip} \hline
\end{tabular}
\end{flushleft}
\end{table*}

The mean values of ${\dot{M}}_{\rm acc}$ over the observed epochs and over several
diagnostics as well as their standard deviations
are presented in
Table~\ref{Table_8}, where estimates from Garcia Lopez et al.\ (\cite{GarciaLopez2006}) and
 Mendigutia et al.\ (\cite{Mendigutia2011a}) are also given for comparison. One can see that, in spite of the
fact that the observations of Garcia Lopez et al.\ (\cite{GarciaLopez2006}) were carried out in 2004, the
discrepancies are within the limits of the standard deviations obtained in both
works.
On the other hand, the results obtained in Mendigutia et al.\ (\cite{Mendigutia2011a}) demonstrate significant
differences with our estimates. The difference is $+0.94$\,dex for HD\,150193 and
$+2.15$\,dex for HD\,190073. Using data on $A_{V}$, $L(H\alpha$), and $L(Br\gamma)$
presented in Mendigutia et al.\ (\cite{Mendigutia2011a}) and from our paper, we estimated contributions of
various factors forming the total discrepancy. These contributions are:
$+0.2-0.3$\,dex follow from differences in fluxes measured on different dates,
$+0.4-0.5$\,dex result from distinctions in $\dot{M}_{\rm acc}$ calibrations (for both
HD\,150193 and HD\,190073). The contributions from differences in adopted stellar parameters
$M_{*}$ and $R_{*}$ are $+0.2$\,dex and $+1.5$\,dex for HD\,150193
and HD\,190073, respectively. Such a significant discrepancy for HD\,190073 is mainly
related to the very large value of the stellar radius, $R_{*}\,=\,8R_{\sun}$, used in
Mendigutia et al.\ (\cite{Mendigutia2011a}). This radius is by a factor of four larger than the value adopted in our work
(Table~\ref{Table_2}).

 We note that the existence of a HAeBe with $T_{\rm eff}$ less than
10\,000 K, $M_{*}=5.1\,M_{\sun}$, and $R_{*}\,=8\,R_{\sun}$ (Mendigutia et al.\ \cite{Mendigutia2011a}) is
not consistent with the PMS evolution model by Palla \& Stahler (\cite{PallaStahler1993}). According to
their model, a PMS object with the initial $M_{*}=5\,M_{\sun}$ starts to become visible at
the birthline with $T_{\rm eff}$ of about 12\,000\,K, and later, closer to the end of the
PMS stage near the main sequence, its effective temperature will become about
16\,000\,K. With the age of about $10^{6}$\,years given for HD\,190073 in
Hubrig et al.\ (\cite{Hubrig2009}) and Mendigutia et al.\ (\cite{Mendigutia2011a}), the star has to be located in the H-R diagram
already near the main
sequence, and with $M_{*}=2-3\,M_{\sun}$ (Table~\ref{Table_2}) its stellar radius
$R_{*}$ cannot be as large as $8\,R_{\sun}$.

 As was discussed in Sect.~\ref{sect:calib_haebes}, the ${\dot{M}}_{\rm acc}$
calibrations derived in Mendigutia et al.\ (\cite{Mendigutia2011a}) lead to very large values of the mass
accretion rates for a considerable part of their targets, which are not in agreement
with the masses of accretion disks ($0.01-0.1\,M_{\sun}$) and the length in time of
the PMS stage for HAes ($10^{5}-10^{6}$\,years). They demonstrate also significant
differences with the results of  Donehew \& Brittain (\cite{DonehewBrittain2011}) and Garcia Lopez et al.\ (\cite{GarciaLopez2006}).

Our ${\dot{M}}_{\rm acc}$ determination is based on:
\begin{itemize}
\item  consistence of values derived from several independent calibrations,
\item agreement of our estimates with those obtained in Garcia Lopez et al.\ (\cite{GarciaLopez2006}), and
\item accordance of our results with the predictions of the PMS evolution model by Palla \& Stahler (\cite{PallaStahler1993}).
\end{itemize}
Clearly, for HAEs with accretion disk masses of the order of $0.01-0.1\,M_{\sun}$ and a
length of the PMS stage of HAEs of $10^{5}-10^{6}$\,years, the expected mean accretion rates
during the PMS stage are of the order of $10^{-6}-10^{-8} M_{\sun}/{\rm yr}$. These expected
rates are similar to the values obtained in our work.

Comparing the accuracy of our estimates with the results obtained by other authors,
we can see that the differences are not significant (see Table~\ref{Table_9}).

\section{Variability of the mass accretion rates}
\label{sect:macc_varia}

Four out of seven of our targets were observed on more than one occasion (from two
to eight). As was discussed in the previous sections, all emission lines chosen as
indicators of ${\dot{M}}_{\rm acc}$ are formed not only in the magnetosphere, but
also in the disk and in the wind, and their contribution to the entire emission is
different for each line. Investigating a temporal behaviour of ${\dot{M}}_{\rm acc}$
using different spectral diagnostics, we have to be certain that a variability
revealed from these indicators relates to a change of ${\dot{M}}_{\rm acc}$
and not to variations of physical and kinematical conditions in the disk and the
wind. For example, the emission in such lines as H$\alpha$, H$\beta$, and DNa~I
originates predominantly  in the wind, while that in the He\,I  line and near the
Balmer jump in the high-temperature region close to the stellar surface
(Pogodin et al.\ \cite{Pogodin2012}). If the variations take place particularly in the outer
circumstellar envelope, they are reflected differently in each of these lines. As a
result, the amplitudes and even the character of the measured ${\dot{M}}_{\rm acc}$
variability derived from different spectral calibrations can be different too. In
such a situation, the revealed changes of ${\dot{M}}_{\rm acc}$ must be considered
as an artifact.

\begin{figure*}
\centering
\includegraphics[width=0.8\textwidth]{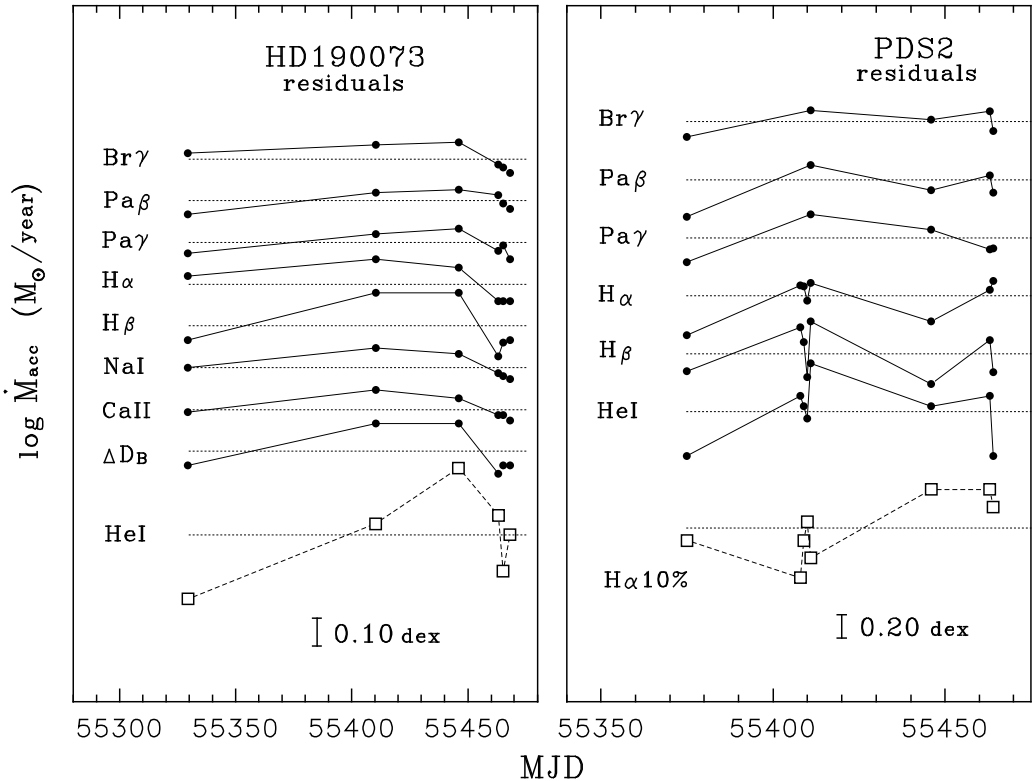}
\caption{
Temporal behaviour of $\log{\dot{M}}_{\rm acc}$ in HD\,190073 and PDS\,2
obtained from different indicators, constructed relative to the mean value for each
indicator.
Thin solid lines join neighbour points for each indicator for better illustration.
The temporal behaviour of the residuals $\log{\dot{M}}_{\rm acc}$ obtained from the
indicator He~I (for HD\,190073, left panel) and from the indicator H$\alpha10\%$
(for PDS\,2, right panel) are marked by open symbols and thin dashed lines. Dotted
lines indicate positions of zero for each indicator. } \label{f11}
\end{figure*}

\subsection{Temporal behaviour of individual accretion diagnostics}
\label{sect:tempor_individ}

Fig.~\ref{f11} illustrates variations of ${\dot{M}}_{\rm acc}$ of
the two targets HD\,190073 and PDS\,2 with the largest number of observations
determined from all
indicators. Since the amplitudes of the variability are smaller than
the discrepancies between the values obtained from different
diagnostics, we  analyzed not the values themselves but the
residuals relative to the mean value for each indicator. As one can
see in Fig.~\ref{f11}, both objects demonstrate variability which:
(a) correlates in a majority of the indicators, and (b) shows
amplitudes of the variations that are very similar for different
diagnostics. It could be intrinsic ${\dot{M}}_{\rm acc}$ variability
but also some geometric effect that affects in the same way our view
of all the different emitting regions.

However, one indicator exists for both objects showing a distinct temporal behaviour
of ${\dot{M}}_{\rm acc}$: it is He~I $\lambda$5876 for HD\,190073 and H$\alpha$10\%
for PDS\,2. The character of the variability of the ${\dot{M}}_{\rm acc}$ value from
the He~I line in the spectrum of HD\,190073 is similar in comparison with other
indicators, but shows a much larger amplitude. In the case of PDS\,2, the
estimations derived from the H$\alpha 10\%$ width strongly differ from all others. A
possible cause for this can be related to the existence of small-amplitude
variations taking place in the circumstellar gas outside the magnetosphere (the
outer disk and the wind) which may influence the full width of the profile but might
be insufficient to have a significant effect on the EW of H$\alpha$. This could
explain a rather small distorting influence of the non-magnetospheric variability
onto the $\dot{M}_{\rm acc}$ derived from the H$\alpha$(L) diagnostic and a rather
significant -- onto that obtained with the H$\alpha 10\%$ indicator.

The other two stars with more than one available spectrum,
HD\,150193 and HD\,135344B, demonstrate the temporal behaviour of
${\dot{M}}_{\rm acc}$ derived from all calibrations that is similar
to the case of HD\,190073. The only exception is the He~I
$\lambda$5876 diagnostic. Variability of ${\dot{M}}_{\rm acc}$
values derived from this indicator is quite different from that
obtained from all other indicators. Presently, no strict model for
the formation of the He~I $\lambda$5876 line in the high-temperature
circumstellar gas around HAeBes exists. We can only assume that a
multi-component variability is likely taking place that is connected
with a change of geometric configuration, optical thickness and
emissivity of the gaseous flows accreted onto the star inside the
magnetosphere. But this has a rather small amplitude and the mean
value derived from He~I differs not so strongly from all the others
estimates (see Fig.~\ref{f10}).

\begin{figure*}
\centering
\includegraphics[width=0.7\textwidth]{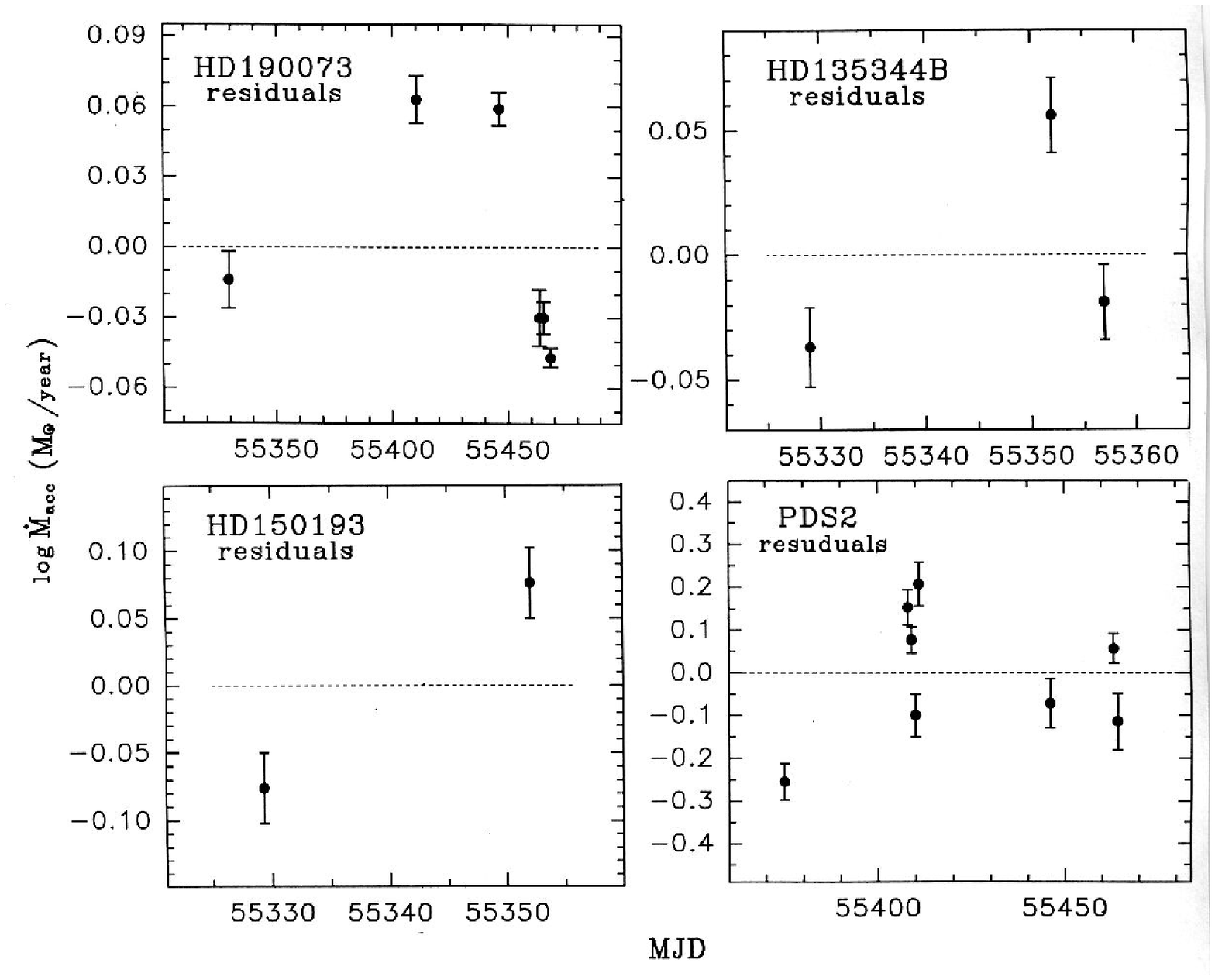}
\caption{ Temporal behaviour of the mean values of the residual
$\log{\dot{M}}_{\rm acc}$ of four targets. The errors at
$\pm1\sigma_{m}$, where $\sigma_{m}$ is the error of the mean, are
marked by bars. The values are obtained using all indicators for
each object (see Fig.~\ref{f8}) except He~I (for HD\,190073,
HD\,135344B, and HD\,150193) and H$\alpha10\%$ (for PDS\,2).}
\label{f12}
\end{figure*}

\subsection{Global accretion variations}
\label{sect:global_variat}

As a measure for the overall accretion variability, Fig.~\ref{f12}
illustrates the temporal behaviour of the mean values of
${\dot{M}}_{\rm acc}$ for the four targets derived from several
diagnostics. The error of the mean value
$\sigma_{m}$=$\sigma$/$\sqrt{n-1}$ (where $\sigma$ is the standard
deviation and $n$ is the number of the used diagnostics) is shown by
vertical bars. Indicators showing a temporal behaviour of
${\dot{M}}_{\rm acc}$ different from all others were excluded from
the calculation of the mean values and their errors (these are He~I
$\lambda$5876 for HD\,190073, HD\,135344B, and HD\,150193, and
H$\alpha 10\%$ for PDS\,2).

HD\,190073 demonstrates variability on the timescale of tens of
days with a spread of the values of about 0.1\,dex with an accuracy at
$1\sigma_{m}$ level of $\pm$0.01\,dex. Short-term  variations (from night to night)
are undoubtedly found in PDS\,2. The observations of HD\,135344B  and HD\,150193
were carried out only on two to three different nights, separated by several to tens
of days. Thus, it is impossible to examine the cause of the variability.

An effective way to analyze the amplitudes and the timescale of accretion
variability was suggested by Nguyen et al.\ (\cite{Nguyen2009}), where a change in amplitude versus the
time interval between observations was constructed for 40 classical TTS. The authors
sampled the timescale range from hours to months and concluded the maximum extent of
the ${\dot{M}}_{\rm acc}$ variability is reached after a few days and amplitudes of
variations are as a rule not more than 0.5\,dex. This result has been confirmed by
other authors. Costigan et al.\ (\cite{Costigan2012}) bolster it with a sample of 10 accreting TTS
observed in the Longterm Accreting Monitoring Project (LAMP) showing that accretion
variation amplitudes do not increase when extending the timescales of the monitoring
beyond a week.

\begin{figure}
\centering
\includegraphics[width=0.45\textwidth]{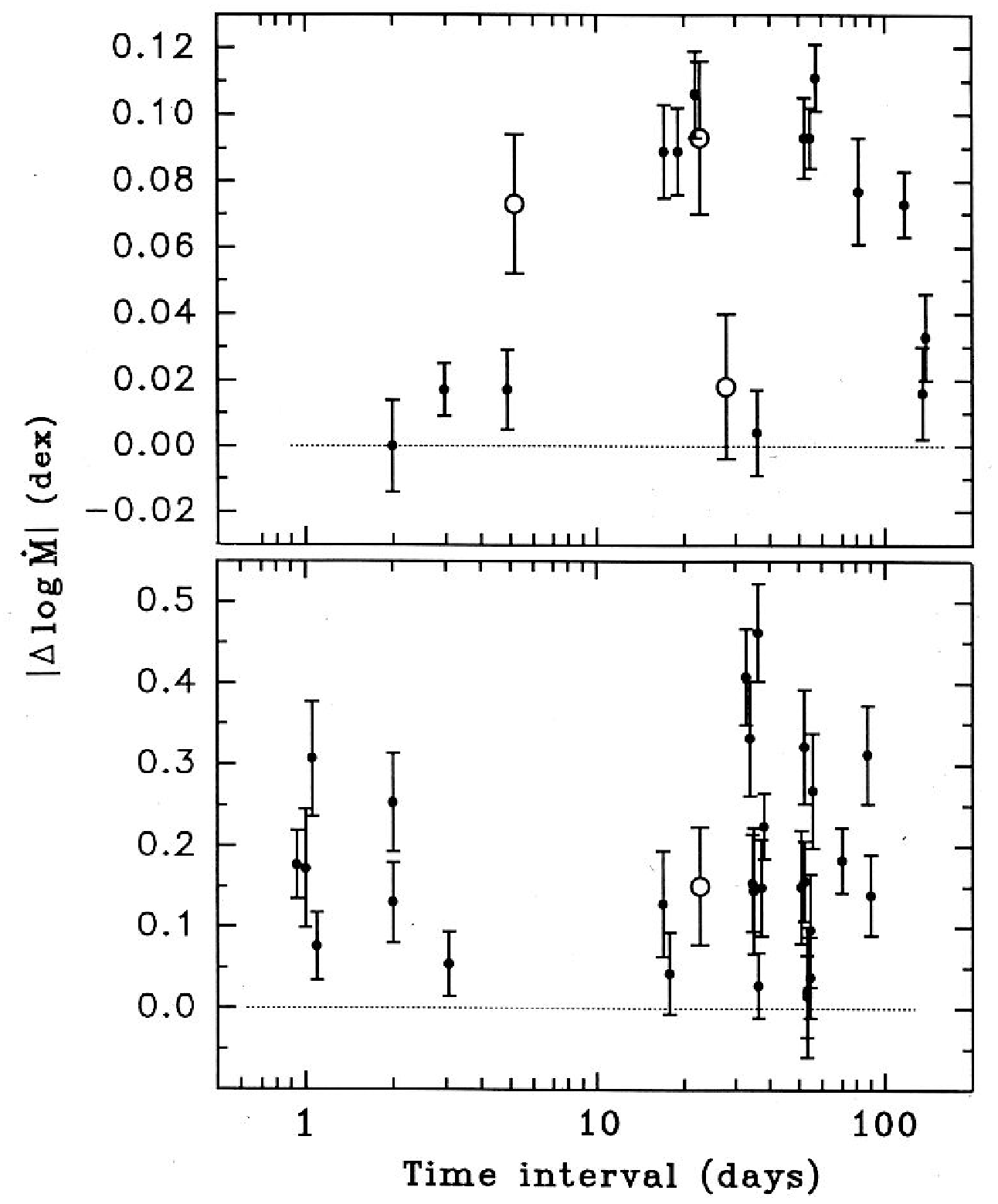}
\caption{Amplitude of variations of the mean $\log{\dot{M}}_{\rm acc}$ of four
targets as a function of time elapsed between the observations. Upper panel: Results
for HD\,190073 (filled symbols) and HD\,135344B (open symbols). Bottom panel:
Results for PDS\,2 (filled symbols) and HD\,150193 (open symbols).} \label{f13}
\end{figure}

In Fig.~\ref{f13} we present our analysis of the ${\dot{M}}_{\rm acc}$ variability
as a function of timescale following Nguyen et al.\ (\cite{Nguyen2009}). Our results are fully in
agreement with those obtained previously by other authors. In the case of PDS\,2,
variations on a timescale of about tens of days  with amplitudes of up to 0.4\,dex
are observed. Besides of that, variability on a timescale near one day is clearly
seen with amplitudes of up to 0.3\,dex. Such a timescale is comparable with
magnetic/rotation periods expected for our targets (Hubrig et al.\ \cite{Hubrig2009}, \cite{Hubrig2011a}).
Therefore variability on such a short timescale is likely to be connected with a
modulation of profiles and intensities of the spectral lines by a rotating
magnetosphere. Using data from Table\,2 ($R_{*}$ and $v$\,sin$\,i$), we estimated
the rotation period of PDS\,2 P = 2.7$\,$sin$i$ days. This object is likely to be
oriented relative to an observer close to ``pole-on'', and its inclination angle $i$
is expected to be of order of $15^{o}$. Such orientation is confirmed by a single
peak emission H$\alpha$ profile (Fig.~\ref{f4}) (see Grinin \& Rostopchina \cite{GrininRostopchina1996}).

Due to the insufficient number of observations, it is not possible
to analyze the ${\dot{M}}_{\rm acc}$ variability on different
timescales for the other three targets. The only conclusion can be
made that a variability has been revealed on timescales from several
days to tens of days. In the case of HD\,190073 we can derive the
rotation period $P$ = 9.1$\,$sin$i$ days using data from Table\,2.
In Hubrig et al.\ (\cite{Hubrig2009}) the upper limit of $i$ is given as
$i\leq40^{o}$. A P\,Cyg-type of the H$\alpha$ profile
(Fig.~\ref{f3}) evidences that this object has to be of an
intermediate orientation relative to an observer (Grinin \& Rostopchina \cite{GrininRostopchina1996}).
With the expected value of $i\sim40^{o}$ we estimate $P$$\sim$6
days. Since no observations separated by short time intervals have
been carried out for this target so far, we cannot examine a
presence of rotational modulations of the line profiles in the
spectrum of HD\,190073.

Another plausible alternative could be that the variability is
the result of stochastic variations in the accretion rate through the disk,
as one might expect if the magneto-rotational instability is the primary source of viscosity
in accretion disks.
More spectral data obtained on different
timescales (from days to tens of days) are needed to rigorously test
the character of the variability of the Herbig Ae/Be stars.

\section{Conclusions}
\label{sect:conclusions}

We present results of observations of a sample of eight magnetic Herbig Ae stars
obtained with the X-shooter spectrographs installed on the 8\,m UT2 at the VLT. This
spectrograph covers simultaneously the whole spectral range from the near-UV to the
near-IR (300--2500\,nm), providing medium resolution, high-quality spectra. We
examined 13 different spectral accretion  diagnostics derived earlier on the basis
of TTS and BD observations on their applicability to HAes. This applicability has
been confirmed for a number of the indicators. The criterion of the applicability
was based on a consistence of the results of ${\dot{M}}_{\rm acc}$ determination one
with another and with the only direct accretion indicator $\Delta D_{B}$,  based
on the CS emission near the Balmer jump. We have used the diagnostics satisfying
the criterion to compute ${\dot{M}}_{\rm acc}$ of all targets, which are ranged from
$2\times10^{-9}$ to $2\times10^{-7}$ $M_{\sun}$ /yr with standard deviations of
0.25\,dex relative to the mean value. The targets in our sample demonstrate a large
variety of spectra. One of the eight targets, HD\,176386, was excluded from our
analysis as an object that has a ``non-developed'' accretion disk. It was not
possible to use all 13 indicators  in all objects. The number of the indicators used
varies from 7 to 10, depending on the target.  An important factor limiting the
application of all indicators is the necessity to re-calibrate a number of empiric
relations constructed earlier for objects showing small ${\dot{M}}_{\rm acc} <
10^{-9}M_{\sun}$/yr. Their application for HAes with higher (${\dot{M}}_{\rm
acc}\sim10^{-7}$$M_{\sun}$/yr) can lead to significant systematic errors. We would
like to emphasize that additional future spectral observations of a larger number of
HAes are needed to improve all calibrations and to expand them to the sample of
earlier-type PMS objects with higher ${\dot{M}}_{\rm acc}$.

An important cause of errors is the uncertainty in determinations of
$M_{*}$ and especially  $R_{*}$ for a number of targets when the
indicators are based on luminosity determination. For objects with
small accretion rate values ($\log\dot{M}_{\rm acc}<-8.5$) errors of
measurements of EWs and widths of emission profiles become an additional
factor for the measurement accuracy.

Our investigation of variability of  ${\dot{M}}_{\rm acc}$ has shown that all four
Herbig Ae stars observed on more than one ocassion demonstrate a change of
${\dot{M}}_{\rm acc}$ of about $0.10\div0.40$\,dex on the timescale of tens of days.
One object observed on a short timescale, PDS\,2, shows also night-to-night
variations with an amplitude of up to $0.30$\,dex. This variability might be related
to the modulation of spectral parameters by a rotating magnetosphere. In the case of
even longer-term variability ($\tau \approx$ tens of days) it remains unclear
whether this variability is connected with a change in the accretion regime or is a
result of rotational modulation.
We note that due to the small number of observations
of targets with determined magnetic/rotation periods, not much can currently be concluded on the
role of magnetic fields in the dynamics of the accretion processes.
Future multi-epoch
observations of magnetic HAes with X-shooter
will be extremely useful to better understand the nature and
variability of $\dot{M}_{\rm acc}$ spectral diagnostics to constrain more realistically
magnetospheric accretion models for Herbig~Ae stars.

{
\acknowledgements

MAP and RVY acknowledge Program N21 of the Presidium RAS and Sci. Scole
N1625.2012.2. }


\label{lastpage}

\end{document}